\documentclass[fleqn,usenatbib]{rasti}

\usepackage{newtxtext,newtxmath}

\usepackage[T1]{fontenc}

\DeclareRobustCommand{\VAN}[3]{#2}
\let\VANthebibliography\thebibliography
\def\thebibliography{\DeclareRobustCommand{\VAN}[3]{##3}\VANthebibliography}

\usepackage{graphicx}	
\usepackage{subfigure}
\usepackage{amsmath}	
\usepackage{algorithm}
\usepackage{algpseudocode}
\usepackage{ulem}
\usepackage{orcidlink}
\usepackage{balance}
\usepackage{hyperref}
\usepackage{xcolor}

\title[Meta-learning for cosmological emulation]{Meta-learning for cosmological emulation: Rapid adaptation to new lensing kernels}

\author[C. MacMahon-Gell\'er et al.]{
Charlie MacMahon-Gell\'er,$^{1}$\thanks{E-mail: c.macmahon@ncl.ac.uk (CMG)}\orcidlink{0000-0001-8944-1977}
C. Danielle Leonard\,\orcidlink{0000-0002-7810-6134},$^{1}$
Philip Bull\,\orcidlink{0000-0001-5668-3101},$^{2,3}$
Markus Michael Rau$^{1}$
\\
$^{1}$School of Mathematics, Statistics and Physics, Newcastle University, Newcastle upon Tyne, NE1 7RU, United Kingdom \\
$^{2}$Jodrell Bank Centre for Astrophysics, University of Manchester, Manchester M13 9PL, UK\\
$^{3}$Department of Physics and Astronomy, University of Western Cape, Cape Town 7535, South Africa
}

\date{Accepted XXX. Received YYY; in original form ZZZ}

\pubyear{\the\year{}}

\begin{document}
\label{firstpage}
\pagerange{\pageref{firstpage}--\pageref{lastpage}}
\maketitle

\begin{abstract}
Theoretical computation of cosmological observables is an intensive process, restricting the speed at which cosmological data can be analysed and cosmological models constrained, and therefore limiting research access to those with high performance computing infrastructure. Whilst the use of machine learning to emulate these computations has been studied, most existing emulators are specialised and not suitable for emulating a wide range of observables with changing physical models. Here, we investigate the Model-Agnostic Meta-Learning algorithm (MAML) for training a cosmological emulator. MAML attempts to train a set of network parameters for rapid fine-tuning to new tasks within some distribution of tasks. Specifically, we consider a simple case where the galaxy sample changes, resulting in a different redshift distribution and lensing kernel. Using MAML, we train a cosmic shear angular power spectrum emulator for rapid adaptation to new redshift distributions with only $O(100)$ fine-tuning samples, whilst not requiring any parametrisation of the redshift distributions. We compare the performance of the MAML emulator to two standard emulators, one pre-trained on a single redshift distribution and the other with no pre-training, both in terms of accuracy on test data, and the constraints produced when using the emulators for cosmological inference. We observe that within an MCMC analysis, the MAML emulator is able to better reproduce the fully-theoretical posterior, achieving a Battacharrya distance from the fully-theoretical posterior in the $S_8$ -- $\Omega_m$ plane of 0.008, compared to 0.038 from the single-task pre-trained emulator and 0.243 for the emulator with no pre-training.
\end{abstract}

\begin{keywords}
Machine Learning -- Cosmological Emulation -- Weak Lensing -- Parameter Inference
\end{keywords}



\section{Introduction}
\label{sec:intro}

Translating the observations of cosmological surveys into constraints on the parameters of cosmological models is a computationally expensive process, requiring us to explore not only the parameter space of said cosmological models, but also of many other 
models for systematic effects, such as, in the case of weak lensing observations, baryonic physics, intrinsic alignment, and photometric redshift errors. The resulting high-dimensional parameter spaces must be exhaustively sampled, often using Bayesian statistical techniques such as Markov Chain Monte Carlo (MCMC) sampling (see, for example, \citealp{padilla2021bayesiancosmology} for a review on MCMC methods in the context of cosmological inference). It may take weeks on many-core HPC systems to achieve converged sampling of the high-dimensional posterior probability distribution.

A large portion of this computational expense results from the requirement to calculate, from theory, the cosmological observables at the sampled coordinates in parameter space. While in isolation such computations are relatively inexpensive (on the order of a few CPU seconds), repeating this process millions of times to obtain the posterior probability distribution becomes a significant computational burden, limiting the speed of research and requiring vast amounts of energy and producing significant amounts of CO$_2$~\citep{To_2023}.

In an attempt to accelerate cosmological inference, many studies have looked at the use of machine learning with neural network (NN) based emulators, as a surrogate for the direct calculations which produce from theory the cosmological observables within an inference pipeline. For example, the \textsc{CosmoPower}~\citep{Spurio_Mancini_2022} emulator speeds up the parameter inference process by up to $O(10^4)$, as compared to direct calculation of the matter power spectrum via numerical solution of differential equations; we will refer to the latter, classical method of calculating the matter power spectrum as being via so-called {\it Boltzmann} codes. Due to the volume of galaxies surveyed, many current and upcoming lensing surveys such as The Vera Rubin Observatory's Legacy Survey of Space and Time (LSST;~\citealp{LSST_SRD}) rely on the measurement of redshift with broad photometric bands (photometric redshifts) to determine the line-of-sight distribution of galaxies. In the context of such surveys, emulators in the style of \textsc{CosmoPower} achieve flexibility to different photometric redshift distributions and systematic models by emulating at the level of the 3D power spectrum. The observables of interest -- typically the real space angular two-point correlation function or its Fourier space equivalent, the angular power spectrum (APS) -- are then obtained by projection of the emulated power spectrum over a given redshift distribution.

\textsc{CONNECT}~\citep{Nygaard_2023_connect} attempts to generalise further and emulate any quantity which can be calculated from the Boltzmann-solver-based code \textsc{CLASS}~\citep{Lesgourgues2011boltzmannclass}, which includes Cosmic Microwave Background power spectra, matter power spectra and more. In order to avoid having to pre-train across all the possible CLASS variables, \textsc{CONNECT} uses an iterative training procedure whereby the model trains alongside the MCMC sampler, learning further from only the high likelihood regions of the parameter space as it progresses and accelerating the chain. Similarly, \cite{Boruah2022} and \cite{boruah2024machinelearninglsst3x2pt} use an iterative training method to create an emulator which, unlike \textsc{CosmoPower}, directly emulates the lensing and clustering auto and cross correlation functions, bypassing the requirement to directly calculate the impact of systematic models and survey redshift distributions on the power spectrum. However, because of this, they note that a new emulator would need to be created for scenarios where the systematic models and galaxy samples differ. A wide range of cosmological emulators based on N-body simulations have also been developed including \textsc{Dark Quest}~\citep{Nishimichi2019}, \textsc{EuclidEmulator2}~\citep{Knabenhans2021}, \textsc{CSST Emulator}~\citep{Chen_2025}, and \textsc{Cosmic Emu}~\citep{moran2023mira}.


In the wider field of machine learning, meta-learning -- in essence training a NN to `learn to learn' -- has proven to be a promising means of creating NNs capable of solving a range of different problems through rapid, inexpensive fine-tuning to new tasks. Meta-learning approaches have been applied in various other fields of physics and seen success in creating adaptable networks capable of training rapidly~\citep{liu2022metaPINN,SALAMANI2023,chen2023pgmtl}, but to our knowledge, no use of meta-learning has been attempted in the field of cosmology to date.

Model-agnostic meta-learning (MAML;~\citealp{finn2017modelagnostic}) is one example of a meta-learning algorithm. Whereas typically neural networks are trained to become extremely proficient at a specific task, MAML seeks to optimise a network to handle a wider range of related tasks. The parameters of the network are optimised to an initialisation which, with very few training data and iterations, can be fine-tuned to solve a novel problem from within the distribution of training tasks. MAML therefore has the potential to create a NN capable of emulating cosmological observables directly and generically. A MAML trained emulator could in principle, with minimal fine-tuning, adapt to new systematic models, gravity theories, or survey samples. Such an emulator could be extremely beneficial not only for reuse within different parameter inference pipelines, but also as a generic emulator for cosmological observables in different contexts such as Fisher forecasting~\citep{EuclidFisher2020}.

MAML training is potentially also highly complementary to the other cosmological emulation efforts mentioned previously. For example, a MAML-trained emulator could be used as a starting point for the iterative training methods of \cite{Nygaard_2023_connect} and \cite{boruah2024machinelearninglsst3x2pt}. Whilst an initial pre-training process would be required, once complete, the MAML based emulator could be substituted in for use with the iterative training methods, potentially leading to faster adaptation to new survey scenarios.

In this study, we will explore the use of the MAML training algorithm in the context of cosmological emulation. We will seek to emulate the cosmic shear angular power spectrum, which describes the strength of the weak gravitational lensing signal across different scales in Fourier space, and is a typical summary statistic in weak lensing surveys. This means that the emulator output can be used to determine the likelihood in an MCMC inference pipeline with no intermediary calculations. In the context of meta-learning, we will train the emulator with the qualitative objective: `\textit{learn to rapidly adapt to different galaxy sample redshift distributions}'.

The galaxy redshift distributions of cosmological surveys are often complex and require many parameters to accurately describe, hence why none of the emulators mentioned previously take as input any parameters describing the redshift distribution. \textsc{CosmoPower} navigates this issue by emulating the 3D power spectrum, which can then be integrated directly over the lensing kernel to obtain the APS. \textsc{CONNECT} and the emulator of \cite{boruah2024machinelearninglsst3x2pt} can emulate quantities which depend on the redshift distribution, such as the APS, but would require re-training for new galaxy samples with different redshift distributions.

Our aim here is to investigate whether an APS emulator trained using the MAML algorithm across a range of different redshift distributions can be quickly and inexpensively fine-tuned for use with an entirely new galaxy sample, without requiring as input any information on the specific survey redshift distribution at hand. In essence, we hope to find some optimal set of network parameters, which minimises the distances to the ideal parameter sets for a range of different redshift distributions. We will then compare the performance of the MAML trained emulator to an emulator trained on a single redshift distribution, and the relative computational cost of each. We will also compare the posteriors obtained using the emulators to a baseline MCMC analysis which computes cosmological observables directly via Boltzmann codes, to ensure the emulators produce accurate constraints.

The rest of this paper will be structured as follows. In Section \ref{sec:MAML_in_cosmo}, we will present a detailed overview of the MAML algorithm, the angular power spectrum, and other important background to this work. In Section \ref{sec:building_training}, we will discuss the design of our emulator, the training process, and the validation and testing of the emulator. In Section \ref{sec:maml_vs_single} we will compare the performance of the MAML emulator to that of an emulator which has been pre-trained on data from a single redshift distribution. To do this we will introduce a novel redshift distribution, fine-tuning each emulator using a few training samples before testing their performance on a larger test sample. Section \ref{sec:maml-vs-fresh} will instead compare the MAML emulator to an untrained emulator, seeking to identify how many training samples would be required to train an emulator from scratch to equal the performance of the MAML emulator on a novel task. In Section \ref{sec:mcmc}, all three emulators will be used within an MCMC analysis to obtain cosmological constraints, and then these constraints compared with a baseline analysis in which the MCMC samples are generated from theory using a Boltzmann code. Finally, Section \ref{sec:conc} will conclude our findings and propose extensions to our work to make the MAML emulator more generalisable and thus more applicable to a wider range of cosmological inference problems.

\section{Meta-Learning in Cosmology}
\label{sec:MAML_in_cosmo}

In this Section we introduce the algorithms for MAML and the \textit{Adam} optimiser. We will also discuss the theory behind the angular power spectrum and its dependence on the redshift distribution, to motivate the investigation of MAML training across a range of redshift distributions.

\subsection{Model-Agnostic Meta-Learning algorithm}
\label{sec:maml}

Optimisation of a NN's parameters, $\Phi$ (typically composed of weights and biases for each node of the network) can generally be expressed as:
\begin{equation}
    \Phi \leftarrow \Phi - \alpha\nabla\mathcal{L}(\mathcal{D},f_{\Phi}),
    \label{eqn:basic_optim}
\end{equation}
i.e. the parameters $\Phi$ are updated by subtracting the gradient of the loss function, $\mathcal{L}$, computed over the training data, $\mathcal{D}$, and the network's prediction given its current parameters, $f_\Phi$ (note that Eq. \ref{eqn:basic_optim} shows the commonly used optimisation method of stochastic gradient descent; SGD). The step-size parameter, $\alpha$, commonly referred to as the learning rate, controls how much the network parameters can change with each step, helping to stabilise the training process. The loss function, $\mathcal{L}$, can take many forms, but a common choice for regression problems such as emulation is to use the mean squared error of the predictions with respect to the training data truths. By iterating Eq. \ref{eqn:basic_optim} many times, one can tune the network parameters to make more accurate predictions by minimising the loss function. (For a more in depth overview on the fundamentals of machine learning see, for example,~\citealp{Murphy2022ProbML}.)

\begin{figure*}
    \centering
    \includegraphics[width=0.8\textwidth]{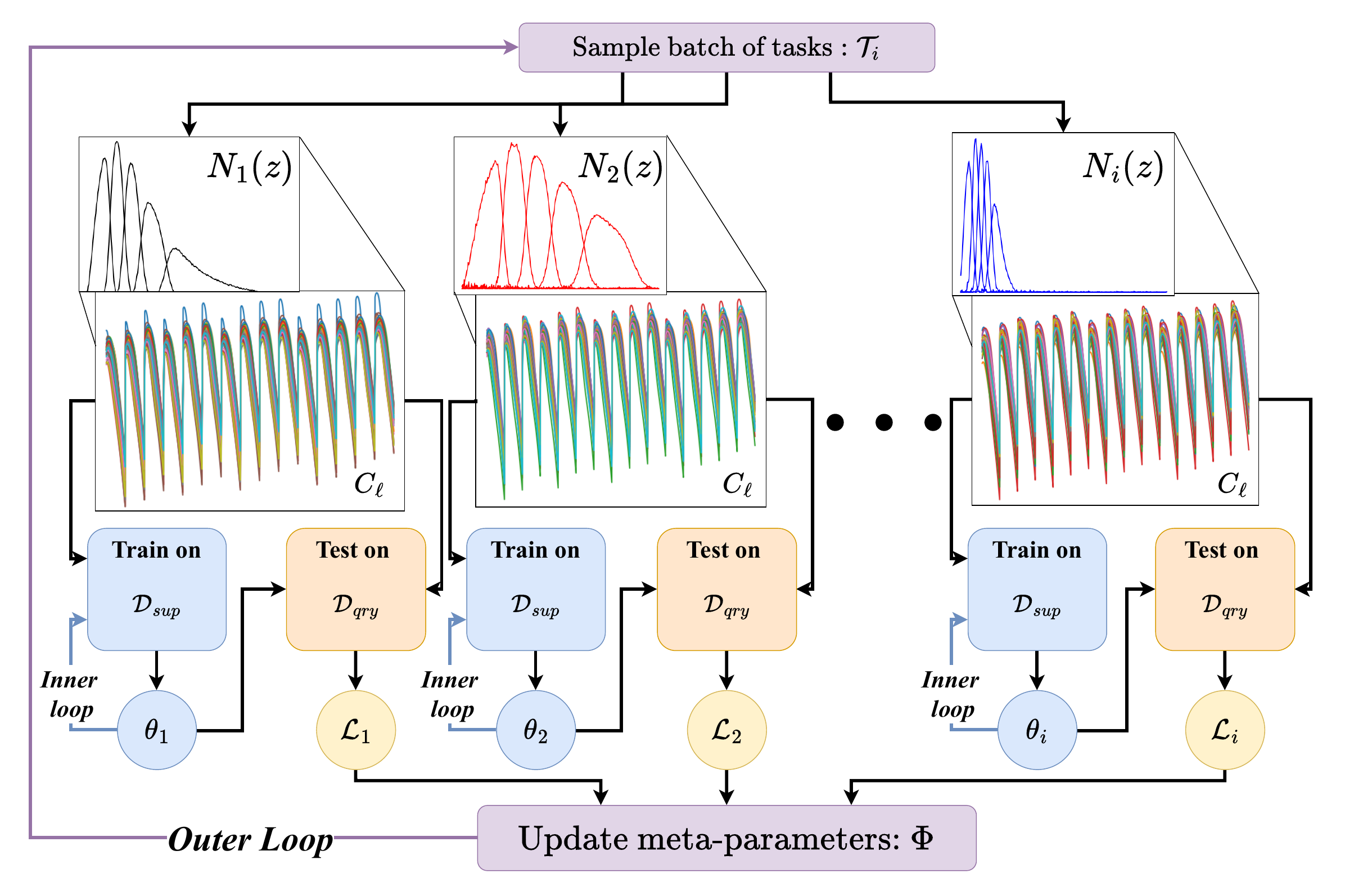}
        \caption{A flowchart of the MAML training process used in this work. Different redshift distributions ($N(z)$) constitute different tasks. At each step in the outer loop, a batch of $N(z)$ are sampled, with APS ($C_\ell$) generated for a range of cosmological parameters for each $N(z)$. The $C_\ell$ are split into support ($\mathcal{D}_{sup}$) and query ($\mathcal{D}_{qry}$) sets. The inner loop trains the NN on $\mathcal{D}_{sup}$ to obtain task-specific parameters, $\theta$, which are then tested with $\mathcal{D}_{qry}$. Once this process has completed for each $N(z)$ in the batch, the query losses ($\mathcal{L}_i$) are averaged together and used to update the meta-parameters of the model, $\Phi$, which will go on to serve as the initialisation for $\theta_i$ in the next batch.}
    \label{fig:maml}
\end{figure*}

Algorithm \ref{alg:maml} shows the MAML algorithm presented in \cite{finn2017modelagnostic}. When training with MAML, we not only want to minimise the loss with respect to a specific task, but across a range of tasks. To achieve this, MAML carries out two optimisation steps. The first optimisation, termed the inner loop, tunes the network parameters for a specific task using a `support' dataset. We denote these task-specific parameters as $\theta$. The loss of these task-specific parameters with respect to an unseen `query' dataset for the same task is then computed. This process is repeated for a batch of $M$ tasks, and the query loss for each task accumulated. Once the entire batch has been processed, the gradient of the accumulated query losses is used to update the meta-parameters of the model, $\Phi$. This is termed the outer loop or meta-optimisation. These updated meta-parameters serve as the new initialisation for fine-tuning the network to a new task. By repeating this process many times, we can reach better initialisations, resulting in better adaptation of the network to new tasks. A diagram of the MAML training process is shown in Figure \ref{fig:maml}.

\begin{algorithm}
    \caption{Model-Agnostic Meta-Learning with Stochastic Gradient Descent}\label{alg:maml}
    \begin{algorithmic}[1]
    \Require $\mathcal{P}(\mathcal{T})$: distribution of different tasks
    \Require $\alpha, \mathcal{A}$: inner and outer learning rates
    \State randomly initialise network parameters $\Phi$
    \While{not done}
        \State Sample batch of $M$ tasks $\mathcal{T}_i$ from $\mathcal{P}(\mathcal{T})$
        \For{all $\mathcal{T}_i$}
            \State Split $\mathcal{T}_i$ into support and query dataset, $\mathcal{D}^{\rm sup}_{i}$ and $\mathcal{D}^{\rm qry}_{i}$
            \State Initialise task-specific parameters $\theta_i \leftarrow \Phi$
            \State Optimise $\theta_i \leftarrow \theta_i-\alpha\nabla\mathcal{L}(\mathcal{D}^{\rm sup}_{i},f_{\theta_{i}})$
        \EndFor
        \State \parbox[t]{\dimexpr\linewidth-\algorithmicindent}{Average the losses on query set over the batch of tasks to get meta-loss: $\mathcal{L}_{meta}=\sum_{T_i}\mathcal{L}   (\mathcal{D}^{\rm qry}_{i},f_{\theta_{i}})/M$}
        \State Update $\Phi = \Phi \leftarrow \mathcal{A}\nabla\mathcal{L}_{\rm meta}$
    \EndWhile
    \end{algorithmic}
\end{algorithm}

It is important to note that the meta-optimisation step in line 10 of Algorithm \ref{alg:maml} requires taking the gradient of the meta-loss, which itself depends on the gradients of the losses over each support set. In common auto-differentiation libraries such as \textsc{PyTorch}~\citep{paszke2019pytorchimperativestylehighperformance} and \textsc{TensorFlow}~\citep{tensorflow2015-whitepaper}, this amounts to a backward pass through the computation graphs for all the task-specific losses. This is highly computationally intensive, particularly in the case where available memory is limited, as is often true when carrying out GPU based computations. However, \cite{finn2017modelagnostic} show that simply omitting the second derivatives and treating the meta-gradient computation as first order results in nearly the same model performance. For an explanation of why this is the case, we refer the reader to~\cite{nichol2018firstordermetalearningalgorithms}.

\subsection{The \textit{Adam} optimiser}

\textit{Adam}~\citep{kingma2017adammethodstochasticoptimization} is a widely used optimisation method for training NNs. It adjusts the learning rate of each network parameter individually by computing first and second moment estimates of the gradients, denoted as $g_t$, which represent the partial derivatives of the loss function with respect to each parameter at time step $t$. These adaptive learning rates help stabilize the training process and often lead to faster convergence to an optimal set of network parameters.

\begin{algorithm}
    \caption{The \textit{Adam} optimisation algorithm. Note that $\beta_i^t$ indicates raising $\beta_i$ to the power $t$. All vector operations and powers are taken elementwise.}\label{alg:adam}
    \begin{algorithmic}[1]
    \Require $\alpha$: initial learning rate
    \Require $\beta_1,\beta_2 \in [0,1)$: Decay rates for moment estimates
    \Require $\epsilon$: Small value to prevent zero division
    \Require $\mathcal{L}(\mathcal{D},f_{\theta})$: Training loss
    \Require $\theta$: Initial model parameters
        \State Initialise 1st moment vector $m_0 \leftarrow 0$
        \State Initialise 2nd moment vector $v_0 \leftarrow 0$
        \State Initialise timestep $t \leftarrow 0$
        \While{not done}
            \State $t \leftarrow t+1$
            \State Get gradients of losses $g_t \leftarrow \nabla\mathcal{L}(\mathcal{D},f_{\theta})$
            \State Update first moment $m_t \leftarrow \beta_1 \cdot m_{t-1}+(1-\beta_1) \cdot g_t$
            \State Update second moment $v_t \leftarrow \beta_2 \cdot v_{t-1}+(1-\beta_2) \cdot g_t^2$
            \State Correct bias for first moment $\hat{m}_t \leftarrow m_t/(1-\beta_1^t)$
            \State Correct bias for second moment $\hat{v}_t \leftarrow v_t/(1-\beta_2^t)$
            \State Update model parameters $\theta \leftarrow \theta - \alpha \cdot \hat{m}_t/(\sqrt{{\hat{v}_t}}+\epsilon)$
        \EndWhile
    \end{algorithmic}
\end{algorithm}

The full \textit{Adam} method from \cite{kingma2017adammethodstochasticoptimization} is shown in Algorithm \ref{alg:adam}, where the \textsc{while} loop indicates training over some arbitrary number of steps, which we shall henceforth refer to as training epochs.\footnote{Lines 9 and 10 in Algorithm \ref{alg:adam} show bias corrections for the moment estimates. These are necessary because in practice, the moments are initialised as vectors of 0's and thus are zero-biased, particularly in the first few training epochs.} In the context of meta-learning and the work presented here, the key thing of note is that parameter updates are dependent on the model's previous performance, by nature of the first and second order moment estimates being moving averages. In essence, the \textit{Adam} optimiser could be considered to hold a `memory' of previous updates through the moment estimates. This can help to balance the update step sizes, reducing the risk of overly large updates in steeper gradient regions and ensuring sufficient movement in flatter areas. This adaptability enhances the optimiser's ability to navigate the loss landscape efficiently.

\subsection{First order MAML with shared \textit{Adam} state}

In the work presented here, we combine the first order MAML (FOMAML) training method presented in Algorithm \ref{alg:maml} with the \textit{Adam} optimisation method presented in Algorithm \ref{alg:adam}. We choose to forgo the calculation of second order gradients in the interest of computational efficiency, as we find our results to be satisfactory with the first order approximation.

Crucially, we also share the \textit{Adam} state (i.e. the latest values of $m_t$, $v_t$, $\beta_1$, and $\beta_2$) between the inner and outer update loops. We find that by sharing the \textit{Adam} state between the task-specific and meta updates, the model converges significantly faster and to a better performing set of meta-parameters than when the inner and outer loops hold unique \textit{Adam} states.

A possible explanation for this improved performance could arise from the accumulation of losses across tasks in line 9 of Algorithm \ref{alg:maml}. By sharing the \textit{Adam} state, the first and second moment estimates encode gradient information from multiple tasks. We therefore not only encode meta-information in the accumulated loss, $\mathcal{L}_{\rm meta}$, but also the outer learning rate, $\mathcal{A}$, potentially leading to more stable meta-updates. In line with this hypothesis, we observed similar fine-tuning performance post meta-training when reusing the same \textit{Adam} optimiser used in the meta-training, and when instantiating a new \textit{Adam} optimiser with a fresh state. This suggests that the shared \textit{Adam} state is primarily useful for the meta-update step, rather than the task-specific inner loop updates.

We caution that while improved performance is seen from sharing the \textit{Adam} state between inner and outer updates in this context, it may not be observed in cases where the tasks differ significantly. When task gradients are poorly aligned, shared moment states could lead to gradient interference, reducing the efficacy of the meta-updates.

\subsection{The cosmic shear angular power spectrum}

Gravitational lensing occurs when light rays pass through space-time distortions caused by the presence of mass, as predicted by Einstein's theory of General Relativity. When light from a distant source galaxy passes through the gravitational potential of a massive object, the change in the light rays' paths causes us to observe an image of the source galaxy with a distorted shape, compared to its true shape.

Weak lensing considers those distortions which are small enough as to only be detectable statistically, through correlating the shapes of many background galaxies. The two most common weak lensing measurements consider either the lensing around particular foreground `lens' galaxies - known as galaxy-galaxy lensing - or correlations in the integrated lensing along the line-of-sight - known as cosmic shear (CS). For a full review on weak lensing see~\cite{Bartelmann_2001}.

We can estimate the cosmic shear signal by correlating the shapes of all pairs of source galaxies in a weak lensing sample. By studying the strength of this correlation for galaxy pairs with different on-sky separations we can observe how the strength of the cosmic shear signal changes with galaxy separation. The observable described here is the two-point angular correlation function of cosmic shear, and its Fourier space equivalent is the angular power spectrum (APS), which we will seek to emulate in this work.

Many surveys (see for example, \citealp{DES,HSC,KiDS}) bin galaxies tomographically by their redshift, allowing us to observe how the cosmic shear signal has evolved through cosmic time. Under the Limber approximation (see, for example, \citealt{Leonard_2023}), the 2D cosmic shear angular power spectrum for a given combination of tomographic bins can be written mathematically as:
\begin{equation}
    C^{GG}_{i,j}(\ell) = \int d\chi \frac{W_{i}(\chi)W_{j}(\chi)}{\chi^2}P_\delta\Bigg(\frac{\ell + 1/2}{\chi}, z(\chi)\Bigg),
    \label{eqn:lensing_aps}
\end{equation}
where the superscript $G$ refers to lensing, $\ell$ is the 2D angular multipole, $\chi$ is the comoving distance along the line of sight which is a function of redshift, $z$. $P_\delta$ refers to the 3D matter power spectrum, which is a function of both on-sky and line-of-sight separation, and $W_i$ and $W_j$ are the lensing efficiency kernels for tomographic bins $i$ and $j$ given by,
\begin{equation}
    W_{i}(\chi) = \frac{3H_{0}^{2}\Omega_{m}}{2c^2}\frac{\chi}{a(\chi)} \int d\chi^{\prime} N_i(z(\chi^{\prime}))
    \frac{dz}{d\chi^\prime}\frac{\chi^\prime-\chi}{\chi^\prime},
    \label{eqn:lensing_kernel}
\end{equation}
where $H_0$ refers to the Hubble constant, $\Omega_m$ is the matter fraction of the universe, $c$ is the speed of light in vacuum, $a(\chi)$ is the scale factor at a given line-of-sight comoving distance $\chi$, and $N(z(\chi^\prime))$ is the redshift distribution of the source galaxies.

\begin{figure*}
    \centering
    \includegraphics[width=\textwidth]{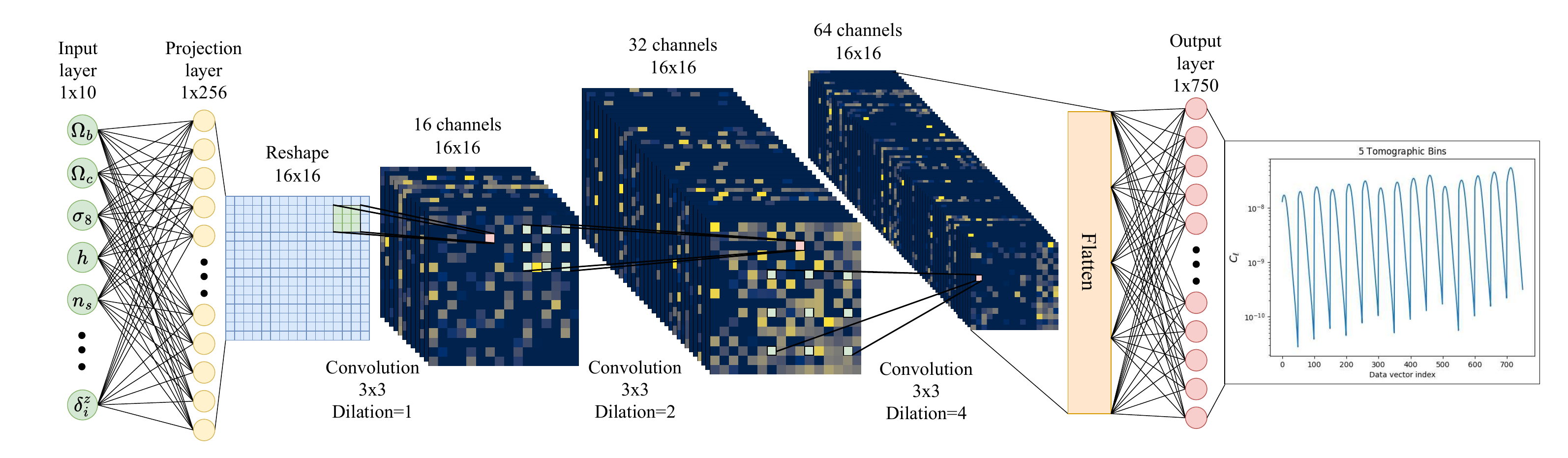}
    \caption{Diagram illustrating the architecture of the neural network used in this work. We take a hybrid approach, combining fully-connected linear layers with convolutional layers in order to capture correlated values in the power spectra data vector as spatial correlations in 2D. Increasing levels of dilation are applied in each layer of the CNN subnet to capture correlations across different scales. The final convolutional outputs are flattened and fed as inputs to a fully connected output layer, which produces the emulated data vector. Images used in the convolutional layers are real activations from the network.}
    \label{fig:nn-diagram}
\end{figure*}

As is apparent from Eqs. \ref{eqn:lensing_aps} and \ref{eqn:lensing_kernel}, $C^{GG}_{i,j}$ has a clear dependence of the redshift distribution of sources, $N(z)$. Additionally, $C^{GG}_{i,j}$ is often computed for the correlations of multiple different tomographic bins. For example, in a survey with 5 tomographic bins, these integrals (implicit in which is the prior computation of $P_\delta$ with a Boltzmann solver) would need to be projected over 15 unique bin combinations at every step of an MCMC analysis, to obtain a set of cosmological constraints. While direct emulation of the matter power spectrum removes the cost of Boltzmann and non-linear modelling, the repeated projection required to obtain all the required APS remains a non-negligible component of the likelihood evaluation, as illustrated in Appendix \ref{app:proj}.

As mentioned previously, real survey distributions are often complex and attempting to parametrise them for use in a NN is not feasible (though methods such as auto-encoders have been proposed to enable this; ~\citealp{ZhangInPrep}). Therefore, in this work, instead of directly encoding information on the redshift distribution as input, we seek to find some meta-parameter initialisation that allows for rapid adaptation to specific distributions, enabling the network to emulate APS with sufficient accuracy that it can be used as a surrogate for Eq. \ref{eqn:lensing_aps} in an MCMC analysis.

It is important to note at this stage that the problem of adapting to different redshift distributions is intended as a basic starting scenario to investigate the applicability of MAML in a cosmological context. To build a versatile emulator capable of adapting to a wide range of novel scenarios, such as different systematic models and gravity theories, would likely require much larger volumes of training data and more rigorous testing which we leave to future work. In this work, we seek to investigate through a simpler problem whether the MAML algorithm is a possible means by which such an emulator could be built. However, in the interest of further motivating future applications of MAML to changing systematic models, we separately perform in Appendix \ref{sec:ia} an exploratory analysis on MAML's ability to adapt to changing intrinsic alignment contaminations.

\section{Building and training the MAML emulator}
\label{sec:building_training}

In this section, we describe the set-up of our MAML emulator, covering the model architecture, training data, and validation of model performance. We use the auto-differentiation library \textsc{PyTorch} to build and train the emulator networks and make the code publicly available~\footnote{\url{https://github.com/CMacM/CosyMAML}}.

\subsection{Emulating angular power spectra in 2D}

In cosmological analyses of probes such as weak lensing, measured quantities are often highly covariant. For example, large-scale overdensities typically correlate with smaller-scale overdensities in the same region due to non-linear hierarchical clustering. Furthermore, cross-correlated APS from different tomographic bins naturally share correlations, making elements of a cosmological data vector inherently covariant.

While such covariances can complicate parameter inference, they can be advantageous for emulation. Inference pipelines typically use a data vector formed by concatenating APS from all unique tomographic bin combinations, as in Fig.~\ref{fig:maml}, where the $C_\ell$ panels show 15 concatenated spectra. A simple fully connected network, such as a multi-layer perceptron (MLP; \citealp{Almeida1996}), may struggle to capture both correlations between spectra and the correlations within each spectrum simultaneously.

To address these challenges, we propose a hybrid approach, illustrated in Figure~\ref{fig:nn-diagram}. The input parameters are first passed through a larger projection layer, which is then reshaped into 2D. This 2D projection is processed by convolutional layers \citep[see][]{oshea2015introductionconvolutionalneuralnetworks}, allowing the network to exploit the strengths of CNNs in learning spatial correlations, and thus capture the correlations expected to be present in the data vector. The output of the final convolutional layer is flattened and passed to a final 1D layer which condenses the result and outputs the emulated data vector.

The architecture in Figure~\ref{fig:nn-diagram} was selected after experimenting with a range of alternatives, including deep multi-layer perceptrons. However, as the focus of this work is to investigate the use of the MAML algorithm in cosmological emulation, we do not provide a detailed comparison of different architectures here, leaving that to future work.

In developing this design, we found that pooling operations \citep{Zafar2022}, which group pixels to capture multi-scale correlations, degraded network performance, likely due to the loss of resolution. Rather than upscaling the latent space and increasing network parameters, we instead use dilated convolutions \citep{yu2016multiscalecontextaggregationdilated} to expand the receptive field without resolution loss. We employ increasing levels of dilation in successive convolutional layers, enabling the network to learn correlations over a range of scales. As shown in Figure~\ref{fig:nn-diagram}, dilated sets of green pixels map to a single pixel in the following layer.

Although not depicted in Figure~\ref{fig:nn-diagram}, dropout \citep{Srivastava2014dropout} is applied within the network. Dropout randomly disables nodes during training, which helps prevent over-fitting and also allows us to estimate prediction uncertainty arising from the network architecture. We use a dropout rate of $0.2$ during both training and fine-tuning, enabling all nodes at test time to use the network’s full capacity for predictions.

We train the emulator with the mean squared error (MSE) loss function for both inner and outer loops. The data vector is log-transformed prior to training, and we standardise both the input parameters and the log-transformed data vector to zero mean and unit variance. This promotes stable training and mitigates issues such as exploding or vanishing gradients that can arise from very large or small values in the network.

\subsection{MAML pre-training data sample}

For the MAML training procedure considered here, we must generate data-vectors with different underlying redshift distributions. We choose two models for the redshift distribution, the first of which is the Smail-type distribution~\citep{Smail1994redshift} often seen in deep weak lensing surveys, which can be written as:
\begin{equation}
    N(z) = z^2 \times \exp\left[-\left(\frac{z}{z_0}\right)^\alpha \right],
    \label{eqn:smail}
\end{equation}
where $z_0$ defines a pivot redshift and $\alpha$ controls the tail of the distribution at high redshift. The second model is simply a Gaussian distribution given by,
\begin{equation}
    N(z) = \exp\left[-\frac{1}{2}\left(\frac{z-z_0}{\sigma}\right)^2\right]
\end{equation}
with free parameters for the mean, $z_0$, and standard deviation, $\sigma$. For each task in our training data samples, we randomly choose either the Smail or Gaussian model.

Each distribution is then split into five bins each containing equal numbers of galaxies (a commonly chosen binning scheme for weak lensing source samples). We model noise in the redshift distributions by adding independent zero-mean Gaussian noise to each redshift bin. For each task, the noise standard deviation is drawn from a uniform distribution between $0$ and $0.1$ and the corresponding noise values are applied discretely to each point in the distribution with a redshift-dependent scaling factor of $(1+z)^{-1}$. The resulting distributions are clipped to ensure $N(z) \geq 0$ and are normalised prior to computing the cosmic shear APS.

The inputs of our emulator constitute five cosmological parameters ($\Omega_c$, the fraction of dark matter in the universe; $\Omega_b$, the fraction of baryonic matter in the universe; $h$, the reduced Hubble constant; $n_s$, the scalar spectral index; $\sigma_8$, the amplitude of matter density fluctuations on $8\, h^{-1}$ Mpc scales) along with shifts on the mean redshift of each tomographic bin, $\delta^i_z$, for a total of 10 inputs. We include shifts in the mean of each tomographic bin to account for the fact that surveys often have redshift measurement uncertainties which must be marginalised over during parameter inference.

\begin{figure*}
    \centering
    \begin{subfigure}
        \centering
        \includegraphics[width=0.45\textwidth]{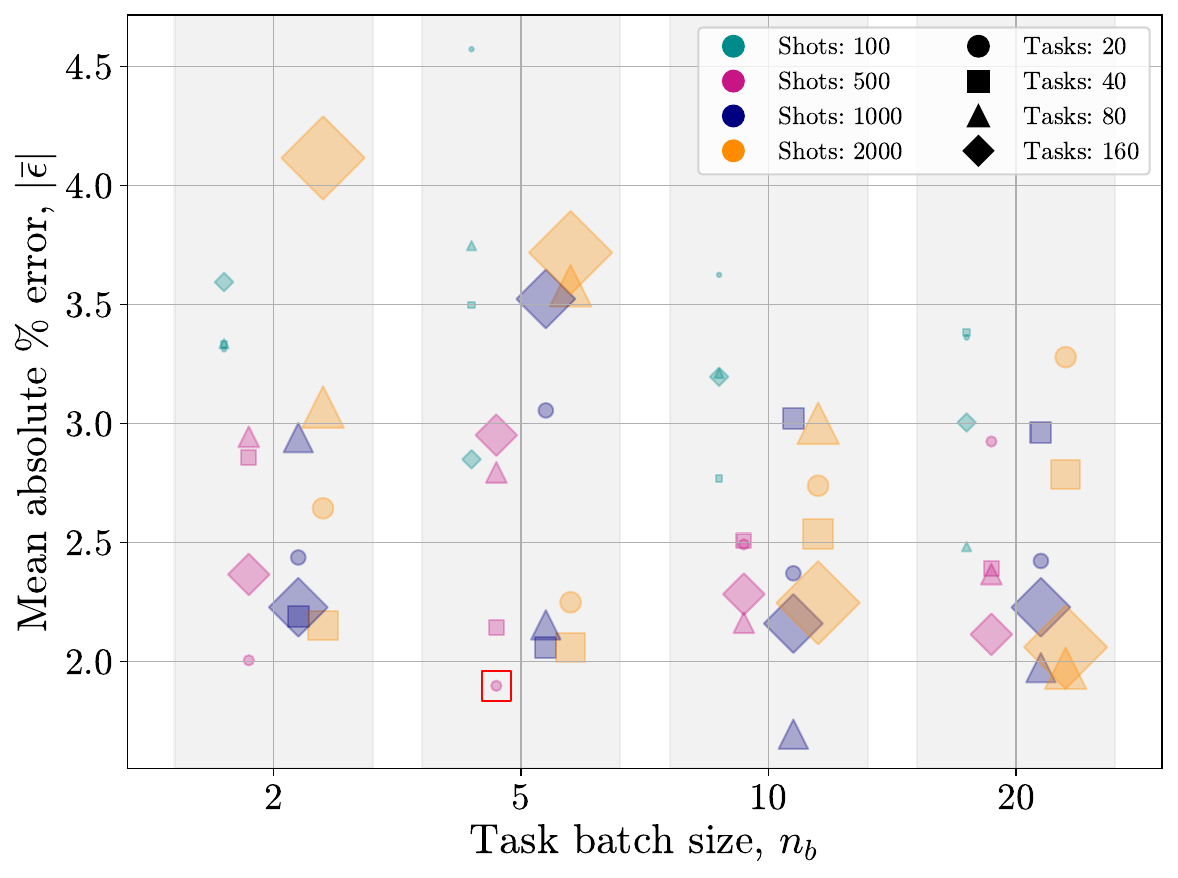}
    \end{subfigure}
    \begin{subfigure}
        \centering
        \includegraphics[width=0.45\textwidth]{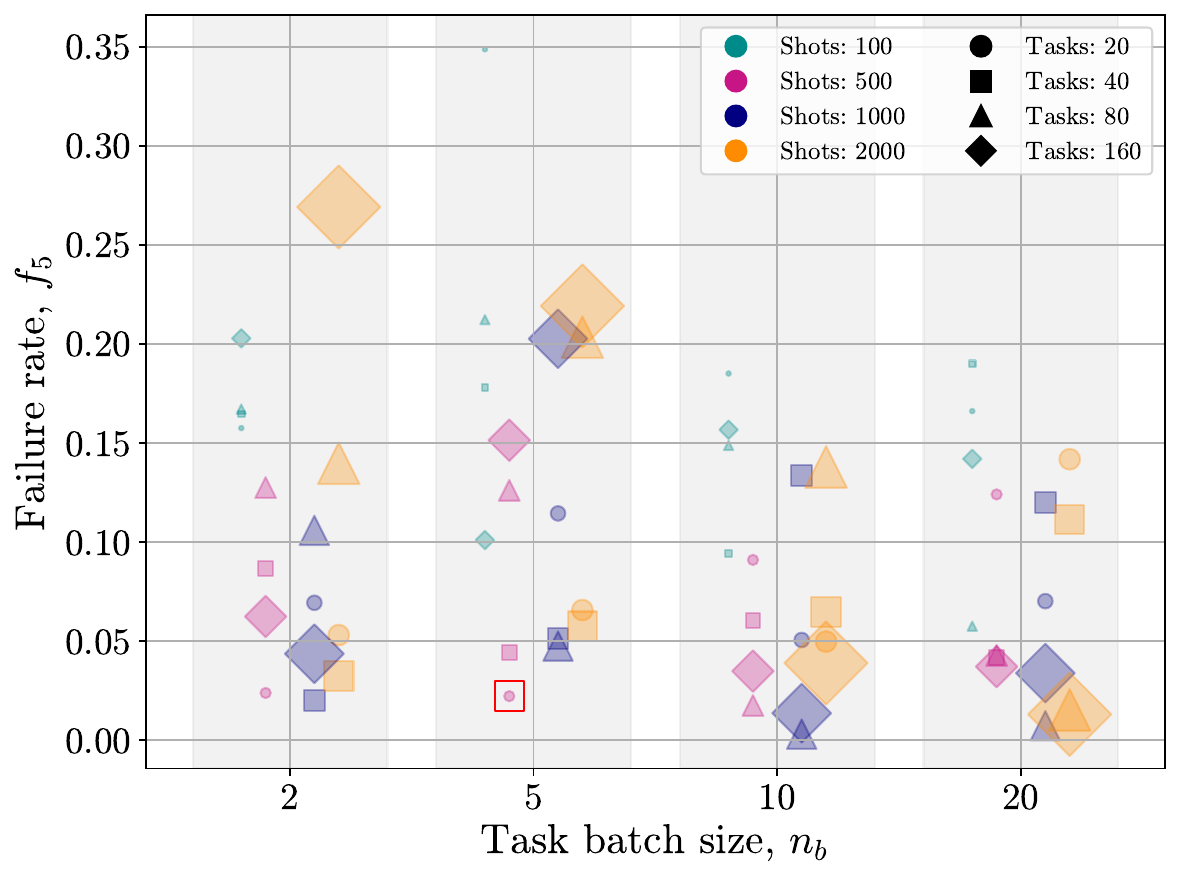}
    \end{subfigure}
    \caption{Mean absolute percentage error (left) and failure rate (right) on the test task, using emulators trained with different numbers of tasks, shots and task batch sizes. The colour of each point is defined by the number of shots, while the marker style represents different numbers of tasks. The size of each marker is determined based upon to the total number of samples required (number of tasks multiplied by number of shots). Groups of points sharing the same task batch size are shown in separate shaded blocks, with the task batch size denoted on the horizontal axis. The point representing the parameters used in this paper is indicated by the red box.}
    \label{fig:data_optim}
\end{figure*}

To ensure adequate coverage of the parameter space, we use a Latin Hypercube sampling~\citep{wei-liem1996lhs}. The range of parameters for the redshift distribution models and for the 10 inputs of our emulator are shown in Table \ref{tab:params}. The cosmological parameter ranges are chosen to be a slightly extended version of the priors used later in Section \ref{sec:mcmc}, and the maximum shift in the mean redshift for each bin is derived from~\cite{LSST_SRD}. The Latin hypercube is re-drawn for each unique redshift distribution.

\begin{table}
    \centering
    \begin{tabular}{cc}
        \hline
        Parameter                           & Training Range  \\ \hline
        \textbf{Emulator Input Parameters}  &                 \\
        $\Omega_c$                          & U(0.165, 0.45)   \\
        $\Omega_b$                          & U(0.025, 0.075) \\
        $h$                                 & U(0.35, 1.15)   \\
        $n_s$                               & U(0.75, 1.15)   \\
        $\sigma_8$                          & U(0.6, 1.05)     \\
        $\delta^i_z$                        & U(-0.0045, 0.0045)  \\ \hline
        \textbf{Smail $N(z)$ parameters}    &                 \\
        $z_0$                               & U(0.1, 0.2)     \\
        $\alpha$                            & U(0.6, 1.0)     \\ \hline
        \textbf{Gaussian $N(z)$ parameters} &                 \\
        $z_0$                               & U(0.2, 1.5)     \\
        $\sigma$                            & U(0.2, 0.6)     \\ \hline
    \end{tabular}
    \caption{Parameter ranges over which data is generated to train the emulator. The emulator takes as input five cosmological parameters and five parameters ($\delta_z^i$) representing shifts in the mean redshift of each tomographic bin. This is intended to enable marginalisation over redshift uncertainty within an inference pipeline. The parameter ranges of the Smail and Gaussian models used to generate the different tasks over which the emulator with MAML train are also shown.}
    \label{tab:params}
\end{table}

In our training sample, cosmic shear data vectors will not only be generated using different combinations of cosmological parameters and tomographic bin shifts (henceforth referred to as input parameters), but also using different underlying $N(z)$. We term each unique $N(z)$ in our training sample a `task' in accordance with the meta-learning nomenclature. For each task, there will be a number of individual data-vectors generated using different input parameter combinations. These individual data-vectors within each task are termed `shots'. For example, a MAML training sample containing a total of $100$ spectra could contain $10$ tasks, with $10$ shots per task, or $5$ tasks, with $20$ shots per task. Such samples could lead to different learning dynamics and performance with MAML, despite containing the same total number of spectra. In order to produce from theory the cosmic shear data vectors needed for training and testing, we make use of the Core Cosmology Library (CCL; ~\citealp{Chisari_2019})~\footnote{\url{https://github.com/LSSTDESC/CCL}}.

\subsection{Determining the optimal number of training samples}

As discussed in Section \ref{sec:maml}, in the meta-update step, loss is accumulated across different tasks. Importantly, the number of tasks over which loss is accumulated in each step is a key hyperparameter (a parameter related to the training of the network or its architecture), which we will call the `task batch size'. Therefore, when attempting to determine the optimal total number of sample spectra to use for MAML training, we have three relevant hyperparameters we can tune: the total number of tasks available for training, $n_\mathcal{T}$, the number of shots, $n_\mathcal{D}$, and the task batch size, $n_b$. Note that in the limit $n_b = n_\mathcal{T}$ the model would train over every available task in each step. This would potentially lead to faster convergence, but at the cost of each update being much slower. It is therefore generally preferable to train over a smaller batch of the available data in each epoch and, with enough epochs, the model will eventually train on all available tasks.

Given computational efficiency is a key focus of this work, we carry out an investigation to determine the optimal number of tasks, shots, and task batch size, as these parameters have a direct influence on the volume of training data required. We do not investigate the impact of changing other hyperparameters. For the MAML algorithm, we use an initial inner learning rate of $\alpha=0.001$ and an initial outer learning rate of $\mathcal{A}=0.01$. For each task, we use $60\%$ of the samples as our support set, $\mathcal{D}_i^{\rm sup}$, with the remainder forming the query set, $\mathcal{D}_i^{\rm qry}$. For the \textit{Adam} optimiser, we use decay rates of $\beta_1 = 0.99$ and $\beta_2 = 0.999$, and $\epsilon = 1\times10^{-8}$.

In order to find the optimal number of tasks, shots, and task batch size, we carry out a grid search with a fixed random seed across 64 combinations of these hyperparameters; using $n_\mathcal{T}$ values of $20,60,100$, and $200$; $n_\mathcal{D}$ values of $100, 500, 1,000$, and $2,000$; and $n_b$ values of $2, 5, 10$, and $20$. For each combination, the emulator is trained subject to an early-stopping criteria. After each meta-update, the emulator is fine-tuned for $64$ epochs on an unseen validation task using set of $100$ data-vectors, and then tested against $4,000$ data-vectors from this same validation task. If the validation test loss does not improve by more than $1\times10^{-4}$ for $20$ consecutive epochs, the meta-parameters are deemed converged and training is terminated. 

In order to quantify the performance of the emulator, we primarily look at two values: the mean absolute percentage error ($|\Bar{\epsilon}|$) across all points in the data-vector and all data-vectors in the test sample (note that we first re-scale the emulator outputs back to the original scale of the data-vector before calculating this), and the failure rate, $f_{5}$, which we define as the fraction of test samples with $|\Bar{\epsilon}| > 5\%$. Therefore, we seek to choose a set of hyperparameters which balances the model performance with the volume of training data required to achieve it.

The results of our grid search with respect to the aforementioned performance metrics are shown in Figure \ref{fig:data_optim}. The metrics are derived from fine-tuning the trained emulator on a new test task (different from the validation task) for 64 epochs with 100 training data-vectors, and subsequently testing the fine-tuned emulator on 4,000 test data-vectors. 

From both metrics, we can infer a few things. Firstly, when a large number of tasks is available, using a larger task batch size leads to better performance. This can be seen from the diamond points representing a total sample of $160$ tasks, which achieve the best performance in general with $n_b = 20$ (though for $n_{\mathcal{D}} = 1,000$ we do see slightly better performance with $n_b = 10$). 

Secondly, using too large a task batch size relative to the number of tasks and shots can also harm performance. This is clearly shown by the points for $n_\mathcal{T} = 20$ and $n_\mathcal{T} = 40$, which often show poorer performance when using task batch sizes greater than $5$. While accumulating loss over larger batches can reduce variance in the meta-update steps and lead to faster convergence, it can also result in over-fitting. It is likely in this case that accumulating loss over too large a portion of the task sample in each meta-update led to this emulator over-fitting to the training sample, and thus when presented with a novel task, it was unable to effectively generalise.

Finally, and most saliently, we see greatly diminishing returns beyond a certain total number of sample data-vectors. To strike a balance between the volume of training data needed and network performance, we opt to use $n_\mathcal{T} = 20, \, n_\mathcal{D} = 500, \, n_b = 5$, resulting in a total of 10,000 training samples. The points corresponding to this choice are indicated by the red boxes in Figure \ref{fig:data_optim}. 

\section{Comparing MAML to standard single-task training}
\label{sec:maml_vs_single}

In this Section, we compare the performance of a MAML-trained emulator on a novel, unseen task with that of an emulator pre-trained using standard methods on a single task. This novel test task is the LSST Year 1 redshift distribution, as defined in~\cite{LSST_SRD}. This comparison reflects a realistic scenario where a standard, non-MAML emulator has been previously trained on a specific galaxy sample, and we desire to repurpose it for use with another galaxy sample or a different selection. More specifically, we seek to identify whether MAML training provides any benefits in this context over more standard approaches.

It is important to note here that the LSST Y1 distribution parameters fall within the range used to generate the MAML training tasks and as such this is considered to be an in-distribution test. We will later investigate the performance of the MAML emulator when adapting to a novel, out-of-distribution task.

\subsection{Pre-training an emulator on a single task}

To pre-train a standard, single-task emulator for comparison with MAML, we use the same total number of samples as for the MAML emulator, but distributed over a single task, such that $n_\mathcal{T} = 1$, $n_\mathcal{D} = 10,000$. We choose a Gaussian redshift distribution for the pre-training task to ensure distinction from the Smail-type LSST Y1 distribution used for testing. We also tested the case where both pre-training and fine-tuning are performed on Smail-type distributions. In this setting, the performance gap between MAML and the single-task emulator narrowed, though MAML retained a slight average advantage and, importantly, showed more stable behaviour across stochastic training realisations. However, because adapting between two very similar task families is a relatively undemanding scenario, we focus here on the more challenging Gaussian-to-Smail case, which better reflects the aim of assessing adaptability across distinct survey conditions.

\begin{figure*}
    \centering
    \begin{subfigure}
        \centering
        \includegraphics[width=0.45\textwidth]{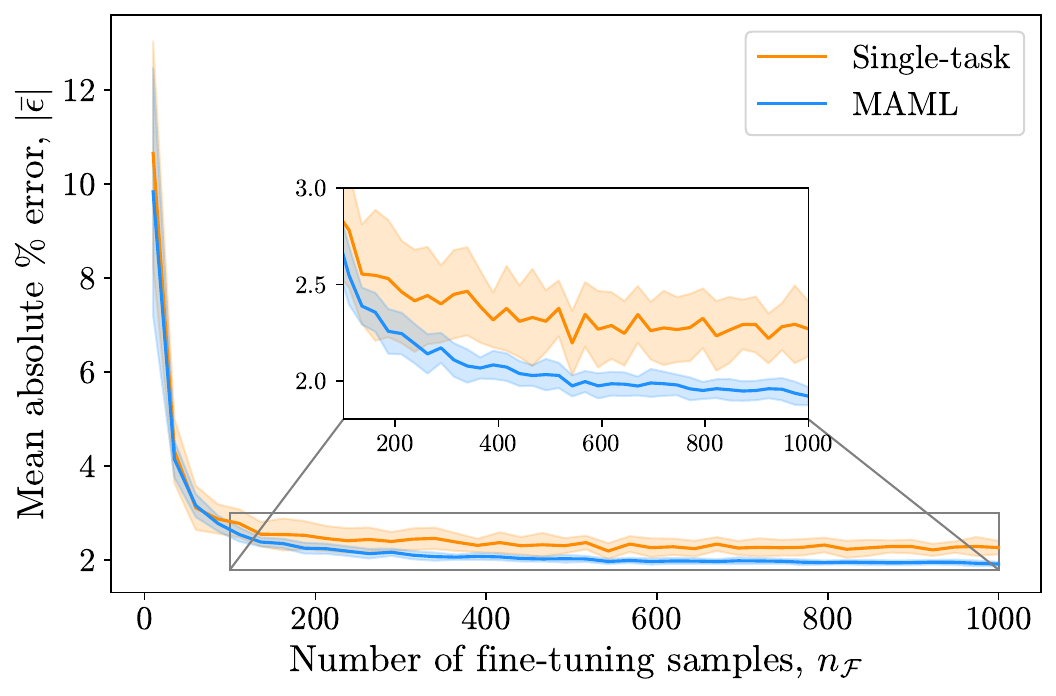}
    \end{subfigure}
    \begin{subfigure}
        \centering
        \includegraphics[width=0.45\textwidth]{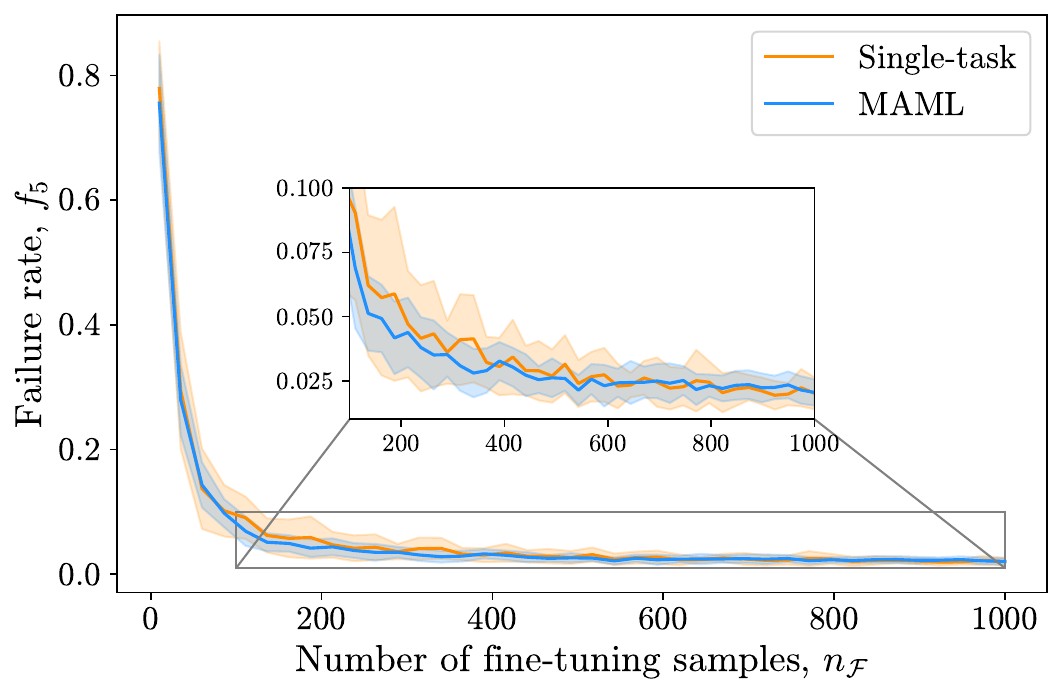}
    \end{subfigure}
    \caption{Mean absolute percentage error (left) and failure rate (right) on the test task, for the single-task (orange) and MAML (blue) trained emulators. The solid lines represent the mean value of each metric over $20$ unique random seeds, and the shaded regions illustrate the $1\sigma$ standard deviation. The insets show a zoomed in view of the results between 100 and 1000 samples. We can see that the MAML emulator achieves lower mean absolute percentage error for all numbers of fine-tuning samples, though both have similar failure rates. Importantly however, the MAML emulator appears to show less deviation in performance with different random seeds, suggesting it is more robust to changing \textsc{cuDNN} optimisations and fine-tuning samples.}
    \label{fig:maml_vs_single}
\end{figure*}

Similar to how in the MAML algorithm we only train on a batch of the total available tasks in each epoch, we train our single-task emulator using batches of 2500 samples randomly selected from the 10,000 available samples. This ensures at each epoch in the training process, the single-task emulator has seen the same number of samples as the MAML emulator (which itself sees 500 samples for each of the 5 tasks selected in each outer epoch).

As in the case of the MAML emulator, we train until the loss on a validation set of 4000 unseen samples from the training task has not improved by more than $1\times10^{-4}$ for 20 consecutive epochs (note that since in this case we are only training on a single task, this single-task emulator does not need to be fine-tuned before being tested on the validation set).

Once the single-task emulator is trained, we fine-tune it, along with the MAML emulator, to a novel task (i.e. emulating spectra derived from an unseen redshift distribution), using different combinations of shot numbers and fine-tuning epochs. When fine-tuning, we reinitialise the \textit{Adam} optimiser for each emulator, such that the pre-training moments do not affect those in the fine-tuning stage. We found this resulted in slightly better performance than re-using the optimisers from the MAML and single-task pre-training.

\subsection{Accounting for stochasticity in machine learning}
\label{sec:rand_gpu}

There are a number of places where stochastic processes can have an impact on the performance of each emulator, potentially leading to an unfair comparison. The first is via the network weight initialisations. Given we use same network architecture in each case, we can easily control for this simply by using the same random seed in both the MAML pre-training and single-task pre-training. This ensures both emulators start with the same weight initialisations, and the same nodes are dropped out in each step of the pre-training process. Of course, given the intention here is to compare two different training processes, the emulators should and will still diverge as pre-training progresses

Of particular importance are stochastic effects in the fine-tuning and testing processes. Perhaps the most obvious source of stochasticity in the emulators' test performances arises from the choice of samples used to fine-tune and test each emulator. In order to control for this, we vary the random seed used to shuffle and split the dataset into training and test samples. 

An important caveat to this approach is that using the same random seed does not guarantee reproducible results when using \textsc{PyTorch}. This is because the \textsc{cuDNN}~\citep{chetlur2014cudnnefficientprimitivesdeep} library used by \textsc{PyTorch} to compile the code to run on a GPU contains optimisation processes that can result in different algorithms being used between different runs, even when the same GPU is used. While it is possible to run \textsc{PyTorch} deterministically by disabling the algorithmic optimisation, this can lead to significantly worse performance. In our case, disabling the algorithmic optimisation led to a factor of 50 increase in runtime on GPU. We can also run deterministically using CPUs, where we observe a comparatively much lower factor of 20 increase in runtime using 48 cores.

However, given the primary aim of this work is to investigate more computationally efficient methods for building cosmological emulators, we argue that it would be inappropriate not to consider the most computationally efficient method of training and running the emulator, which is of course using GPUs. Therefore, if an emulator's performance is significantly impacted by the \textsc{cuDNN} optimisation routines, this should be taken into consideration alongside the variation in performance due to the choice of training samples. As such, we do not force \textsc{cuDNN} to run deterministically when varying the seed, meaning any variation in performance observed will be not only due to do differences in the fine-tuning and test data, but also algorithmic choices made by \textsc{cuDNN}.

\subsection{Performance metrics comparison}

Each emulator is fine-tuned to the novel task using 40 different numbers of fine-tuning samples, uniformly spaced between 10 and 1,000, in order to investigate the difference in performance between the two emulators across a range of fine-tuning sample sizes. We chose to train for a set number of 64 epochs (determined by visual inspection of the training loss) rather than checking convergence on a validation set. This is because in testing we found the convergence check to be unreliable when using a validation set of only $O(10)$ to $O(100)$ samples.

Once fine-tuned, the emulators are then tested on 20,000 unseen samples. \footnote{It should be noted that the input parameter space for the novel task is sampled via a Latin hypercube sampling of 30,000 points. Therefore, each fine-tuning sample and test sample are drawn from within this larger hypercube. We tested the impact of drawing new Latin hypercubes for each number of fine-tuning shots to ensure coverage of the parameter space, but found no significant difference in emulator performance compared to simply sub-sampling from the larger hypercube, likely indicating the hypercubes are over-sampled relative to their dimensionality.}
For each fine-tuning sample size, we test 20 different random seeds to obtain an estimate of the mean and variance on $|\Bar{\epsilon}|$ and $f_{5}$ with respect to the chosen fine-tuning samples and varying \textsc{cuDNN} optimisations.

Figure \ref{fig:maml_vs_single} shows the results of this test. We can see the MAML emulator on average outperforms the single-task emulator in terms of $|\Bar{\epsilon}|$, though the difference is perhaps less significant than one might initially expect. Given that both emulators achieve similar values of $f_{5}$, it suggests that the lower value of $|\Bar{\epsilon}|$ is the result of reduced scatter in the MAML emulator's predictions. We will go on to see later how this performance translates in the context of cosmological parameter inference, but these results suggest that a cosmic shear angular power spectrum emulator trained for a single galaxy sample / redshift distribution, can be retrained with relatively little expense for application to a novel sample and still achieve good performance. Interestingly, the MAML emulator also exhibits more consistent performance with respect to different random seeds, with the $1\sigma$ deviation in performance for each fine-tuning sample number being narrower than that of the single-task emulator on both $|\Bar{\epsilon}|$ and $f_{5}$.

Again, it is important to state that both emulators have pre-trained on the same total number of spectra and subsequently been fine-tuned and tested using the same samples, but by splitting these spectra across multiple redshift distributions and using the MAML training algorithm, the MAML emulator is able to achieve better, more consistent performance.

We also performed this same test using different numbers of fine-tuning epochs, from 32 up to 512. While 32 epochs showed slightly diminished performance in both emulators, increasing the number of epochs beyond 64 did not show any significant benefit. We also did not observe any significant difference in the relative performance of the emulators, and so we do not include these tests here.

\subsection{Computational requirements of MAML and single-task training}
\label{subsec:compreq}

Given that a key motivation of this work is related to the computational burden of cosmological inference and training cosmological emulators, it is important that we consider the relative pre-training time for each emulator. Since the MAML algorithm contains two optimisation loops, it is naturally slower than training an emulator in the standard fashion on a single task. The hardware used for these tests constituted a single Nvidia(R) A40 GPU and two Intel(R) Xeon(R) Gold 5220R CPUs, with a combined core count of $48$. Note that when we later refer to CPU core-hours, we are considering the time required when utilising a single core on this model of CPU and when we refer to GPU hours, we are considering time required when fully utilising the A40 GPU. We choose this nomenclature in order to simplify the interpretation of the wall times presented.

Once pre-trained, both emulators require the same amount of time to fine-tune and make predictions; we do not account for this in our comparison. Fine-tuning for such few epochs on such few samples is so rapid ($O(10^{-5})$ GPU hours or $O(10^{-3})$ CPU core-hours) that it is effectively negligible in comparison to the time required to generate pre-training data-vectors or perform an MCMC analysis.

Both emulators are trained using a total of 14,000 sample data-vectors, which require $11.2$ CPU core-hours to generate (note that the cost of generating multiple redshift distributions for the MAML sample is negligible). For the single-task emulator, pre-training takes approximately $0.014$ GPU hours, or $10$ CPU core-hours. For MAML, this increases to $0.044$ GPU hours, or $30$ CPU core-hours. Clearly, MAML training presents a significant overhead, tripling the time required to pre-train the emulator. One might initially expect the overhead should only double the training time, as there is only one extra training loop, however, extra time is required since the inner loop performs 5 iterations before each meta-update, and the convergence check for the MAML emulator requires a small fine-tuning step on the unseen task before the validation loss can be estimated.

The picture is somewhat changed if we also include the time required to generate the training data-vectors. Including this time, the difference between training the single-task and MAML emulator becomes almost negligible when a GPU is available. Without the availability of a GPU, MAML training (including time to generate training data) overall takes roughly twice as long as training for a single task. Therefore, whether MAML training is worthwhile in this case largely depends on the computational resources available, and how well the emulator is required to perform for the given task. In the context of pre-training an emulator for generalisability, re-application to different survey samples and for use by researchers without access to high-performance computing facilities, it is likely those who are pre-training the emulator will have access to GPU hardware. As such, MAML training appears to provide a noticeable benefit in accuracy for very little overall increase in training time.
\begin{figure*}
    \centering
    \begin{subfigure}
        \centering
        \includegraphics[width=0.45\textwidth]{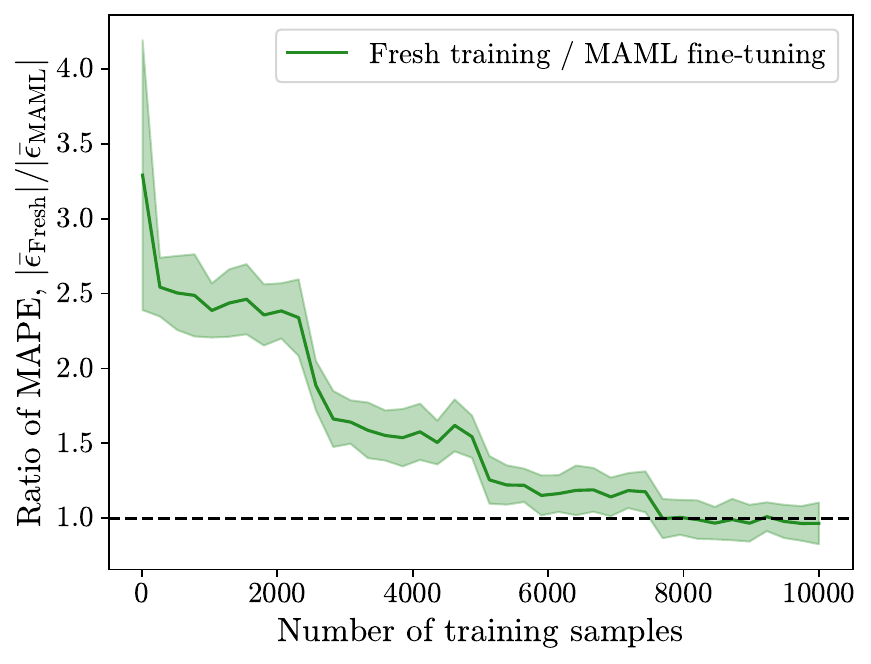}
    \end{subfigure}
    \begin{subfigure}
        \centering
        \includegraphics[width=0.45\textwidth]{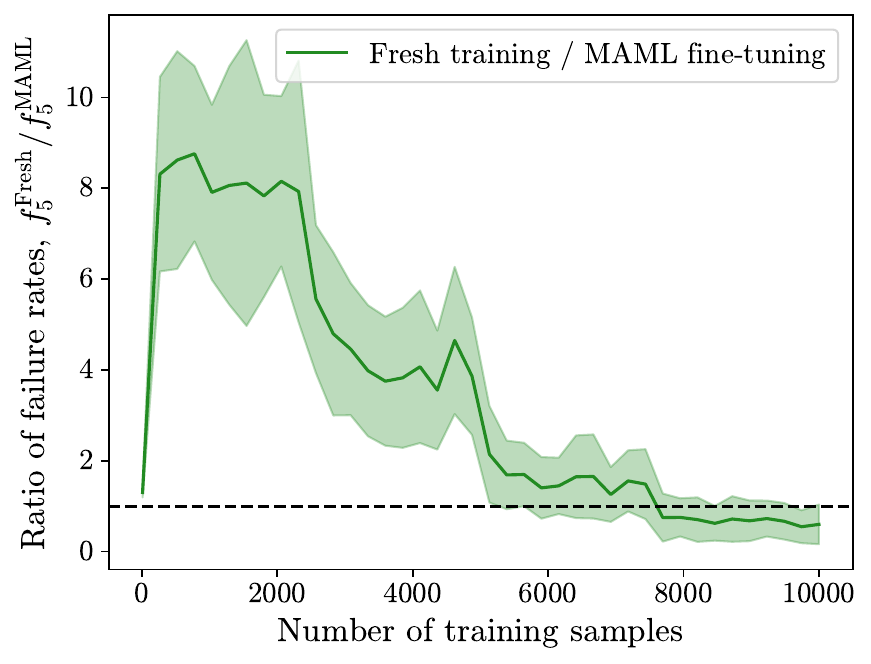}
    \end{subfigure}
    \caption{Ratio of mean absolute percentage error (left) and failure rate (right) on the test task for the fresh emulator with respect to the MAML emulator. The horizontal axis shows increasing numbers of training samples used to train the fresh emulator, while the number of samples given to fine-tune the MAML emulator remains fixed at $100$. We see that the fresh emulator starts to match or exceed the performance of the MAML emulator once more than about $8,000$ samples are provided for training. The solid lines show the average ratio for $20$ different selections of training and test data, while the shaded region indicates the $1\sigma$ deviation in the ratios across these $20$ selections.}
    \label{fig:maml_vs_fresh_id}
\end{figure*}
\section{Comparing MAML to an untrained emulator}
\label{sec:maml-vs-fresh}

In this section, we will seek to compare the performance of our pre-trained MAML emulator to that of an emulator which is trained from scratch for the novel task. The intention here is to investigate how much training data and compute time would be required to build an emulator from scratch for each new task, and thus determine whether building a MAML emulator for fine-tuning to new problems shows any real benefit over simply constructing new emulators for each task.

\subsection{Baseline comparison: In-distribution tasks}

To carry out this comparison, we fix the MAML emulator fine-tuning parameters to 64 epochs and 100 samples. We choose 100 samples as we are interested in fine-tuning with as few samples as possible and beyond this point we see diminishing returns in performance from Figure \ref{fig:maml_vs_single}. We then train a new emulator with no prior training (henceforth referred to as the `fresh' emulator) on the LSST Y1 test task, once again using the random seed to initialise the model. We test 40 different numbers of training samples for the fresh emulator uniformly distributed between 10 and 10,000 (note that where the number of training samples for the fresh emulator exceeds the number of fine-tuning samples for the MAML emulator, we use the first 100 samples of the training set for fine-tuning). Given here we are considering a case where an emulator is built from scratch with limited data to compare to a fine-tuned MAML emulator, we do not implement a convergence check using a validation sample and instead train for 64 epochs in all cases, as no validation sample was used for fine-tuning the MAML emulator or single-task emulator in the previous section. 

Once trained, the fresh emulator and fine-tuned MAML emulator are tested on the remaining samples of the 30,000 sample data-set. We repeat this process for 20 random seeds (note that we only vary the seeds used to select the training, validation, and test data; the seed controlling the network weight initialisations and layer dropouts remains fixed). As the MAML emulator was trained using a batch of 5 tasks with 500 samples per task, we also train the fresh emulator with a max batch size of 2,500. In cases where the total number of training samples is less than this value, we simply train on all samples simultaneously.

The results of this test are shown in Figure \ref{fig:maml_vs_fresh_id}, where we plot the ratio between the performance of the fresh emulator and the MAML emulator. We see that the fresh emulator underperforms the MAML emulator, until a training sample size of 8,000 samples is reached. With increasing numbers of training samples from this point, the fresh emulator is able to equal or better the MAML emulator's performance. 

Comparing this value to the total number of 10,100 samples used to pre-train and fine-tune the MAML emulator, this strongly suggests MAML training is worth considering in cases where an emulator is expected to be re-used for different galaxy samples (such as for various sub-samples within a survey), as the number of training samples generated need only be increased by 25\%.

\subsection{Out-of-distribution comparison: Multi-modal Gaussian distribution}

\begin{figure*}
    \centering
    \begin{subfigure}
        \centering
        \includegraphics[width=0.45\textwidth]{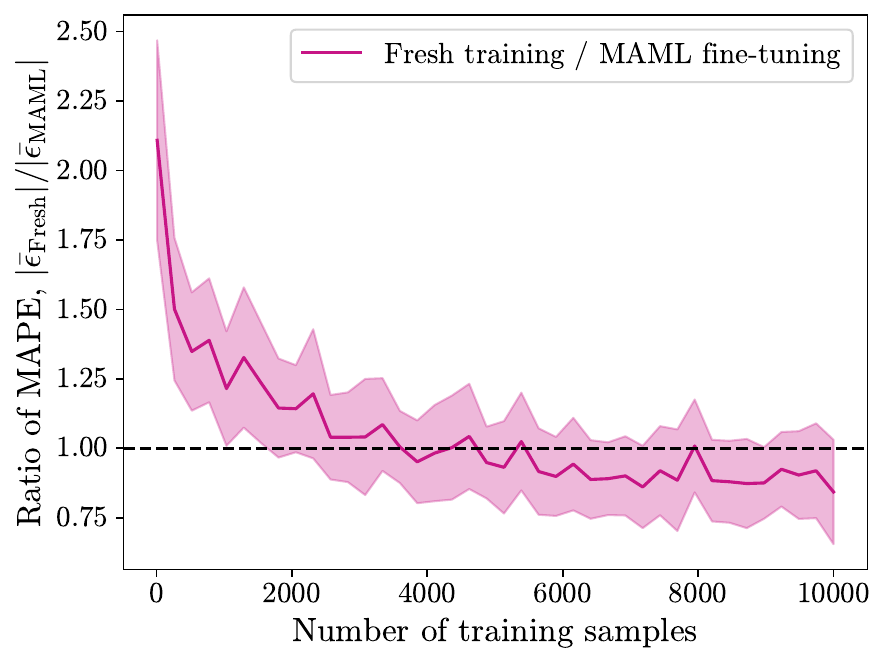}
    \end{subfigure}
    \begin{subfigure}
        \centering
        \includegraphics[width=0.45\textwidth]{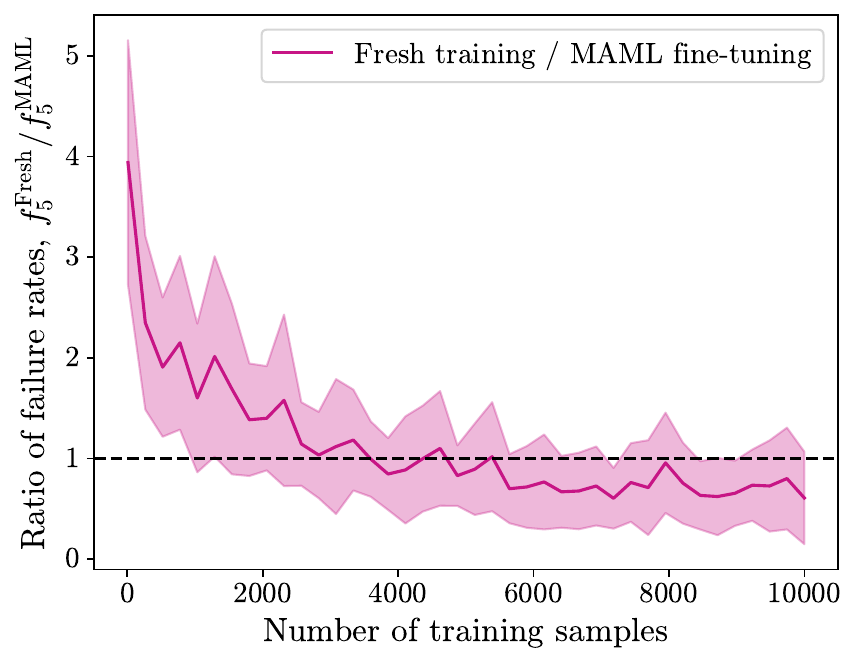}
    \end{subfigure}
    \caption{Ratio of mean absolute percentage error (left) and failure rate (right) for the fresh emulator with respect to the MAML emulator. In this case, a multi-modal Gaussian distribution is used for the test task, to test how well the MAML emulator adapts to out-of-distribution tasks. The horizontal axis shows increasing numbers of training samples used to train the fresh emulator, while the number of samples given to fine-tune the MAML emulator remains fixed at 100. We see that the fresh emulator starts to roughly match the performance of the MAML emulator once more than 4,000 samples are provided for training. As in the previous plots, the solid lines show the mean ratio for 20 different selections of training and test data, while the shaded region indicates the $1\sigma$ deviation in the ratios across these 20 selections.}
    \label{fig:maml_vs_fresh_multipeak}
\end{figure*}

An interesting point of comparison to a freshly trained emulator is the case of out-of-distribution tasks, i.e. redshift distributions with parametrisations outside the range of tasks over which the MAML emulator was pre-trained.

In order to make this test particularly challenging, we consider a novel task redshift distribution that is multi-modal, constructed via the combination of three distinct Gaussian distributions. We note that the key challenge here is the multi-modality itself; whether the peaks arise from Gaussians or from mixing Gaussian and Smail distributions would only make minor quantitative differences. This is because the lensing kernel integrates over the full redshift distribution, so the dominant factor for emulator performance is the presence of multiple separated peaks rather than the exact analytic form of each component.

Whilst such an extreme distribution is unlikely to be observed in reality, the intention of this comparison is simply to assess how well the MAML emulator can generalise to a task well outside the distribution of tasks on which it was pre-trained.

The results of this test are shown in Figure \ref{fig:maml_vs_fresh_multipeak}. We see that, in this case, the freshly trained emulator outperforms the MAML emulator after only 4,000 training samples. This suggests that the meta-parameters obtained through the MAML training are most suitable for tasks within the range of those over which the emulator was trained. However, given the freshly trained emulator still requires more training samples than are required by MAML fine-tuning to achieve the same performance, it does indicate that some of the features learned by the MAML emulator in its pre-training are applicable to the out-of-distribution task.

To further support this, better performance on the out-of-distribution task can be obtained from the MAML emulator by providing more fine-tuning samples. We do not show a comparison of this here, but as a general trend, we observed that to achieve the same performance between a freshly trained emulator and a fine-tuned MAML emulator on the out-of-distribution task, the fresh emulator required 40 times as many training samples, compared to 80 times as many for an in-distribution task. Whether this means the emulator has truly `learnt to learn' or has simply built a robust set of features for changing redshift distributions is an avenue of investigation which we leave for future work.

\section{Comparing emulators within an MCMC analysis}
\label{sec:mcmc}

\begin{figure*}
    \centering
    \includegraphics[width=\textwidth]{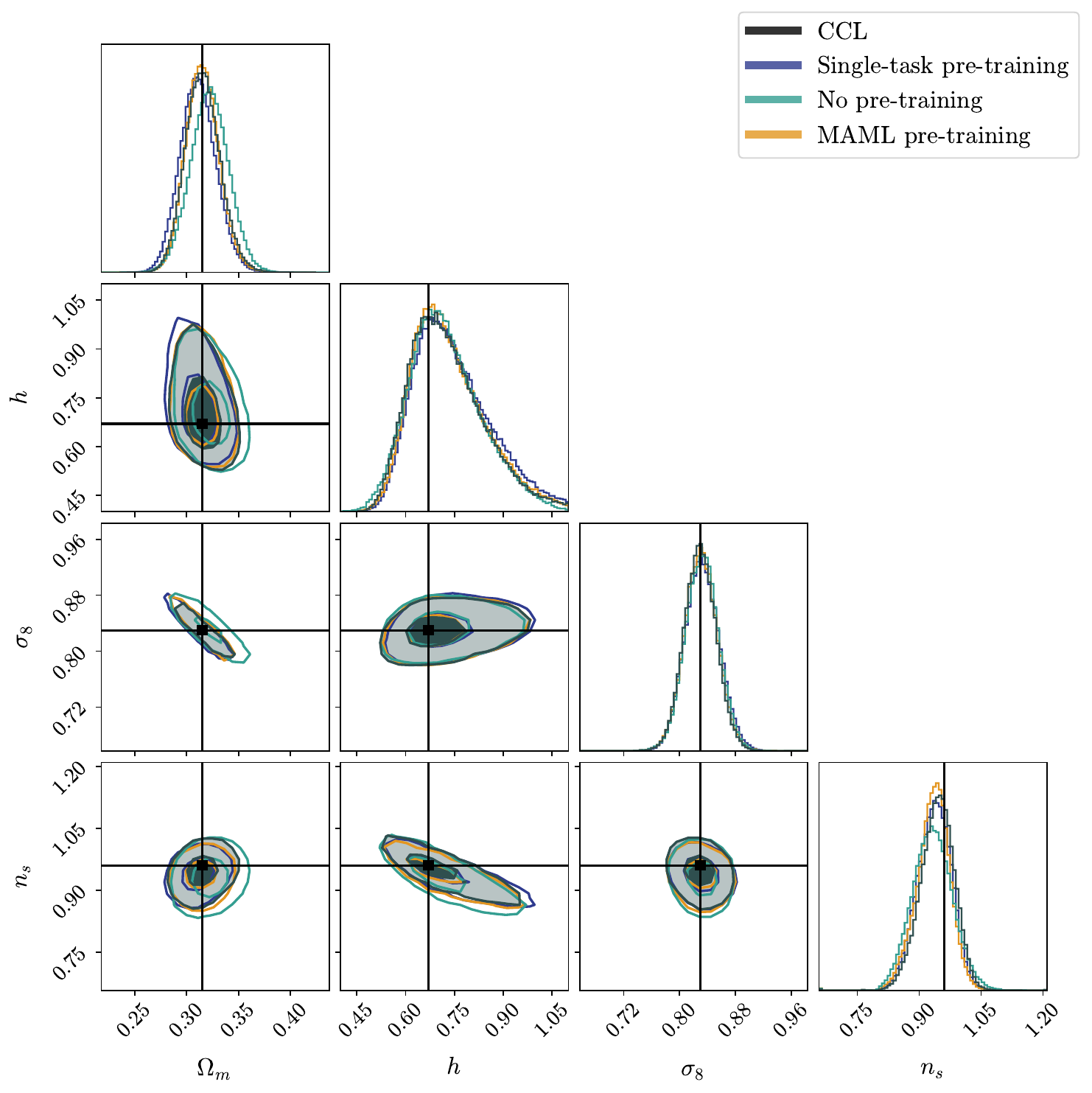}
    \caption{Parameter constraints from all MCMC chains. The black contours are obtained by using CCL in the likelihood function to compute the angular power spectra from theory. The blue, green, and orange contours are obtained using the single-task, fresh, and MAML emulators respectively, with the fiducial cosmological parameters indicated by the black squares. The fresh emulator shows a small bias from the CCL constraints while both the MAML and single-task emulators recover all constraints well.}
    \label{fig:mcmc_contours}
\end{figure*}

Thus far, we have looked at specific metrics to measure the accuracy of emulators trained in 3 different ways. However, given a primary objective of creating such emulators is to accelerate cosmological inference pipelines, it is important that we compare their performance in terms of the emulated posteriors obtained using an inference technique such as MCMC, as quantifying the error on emulated test data-vectors does not alone tell us how well an emulator will perform when used for parameter inference.

To provide a baseline point of comparison, we run a full MCMC analysis using CCL to generate from theory the spectra for each sample in the chain. The cosmological parameter values which define the fiducial cosmology used here are shown in Table \ref{tab:mcmc_values_priors}, along with corresponding priors. The redshift distribution used for the MCMC analysis is the same LSST Y1 distribution as used for the novel task considered in Section \ref{sec:maml_vs_single}. To obtain a covariance matrix associated with our fiducial data-vector (which itself is generated using CCL and our fiducial parameters) we use \textsc{TJPCov}~\footnote{\url{https://github.com/LSSTDESC/TJPCov}} to construct a matrix representative of an LSST Y1 like sample, with a galaxy number density of $N_{\rm gal} = 10\, \text{arcmin}^{-2}$ and shape noise $\sigma_e = 0.26$.

To sample the posterior distribution, we make use of the \textsc{emcee}~\citep{Foreman_Mackey_2013} package. For both the baseline CCL inference and the emulator-based inferences, we use 76 walkers initialised in a Gaussian ball with a 10\% spread around the fiducial parameter values, and check convergence every 1,000 steps by computing the autocorrelation time for each parameter. We consider the chains converged when the total chain length exceeds 50 times the estimated autocorrelation time for all parameters. 

In the CCL-based MCMC analysis with a theoretical likelihood, convergence was reached after 15,000 steps of the chain. For the hardware used in this work, this equates to approximately 1,225 CPU core-hours. Due to the simpler parameter space we are considering here, this is much lower than typical cosmological inference problems. For example, the 24-dimensional MCMC sampling done in \citealp{mill2025taking} takes around 9,000 CPU core-hours to converge.

For the emulator-based MCMCs, we also use the \textsc{emcee} package, but additionally consider the case where a GPU is used to produce the emulated power spectra in the likelihood function. Convergence in the emulated chains is achieved in 17,000, 18,000, and 20,000 steps for the single-task, MAML and fresh, which all equate to approximately 300 CPU core-hours or 1 GPU hour. The variation in steps required for convergence is the result of both stochasticity in the sampling process and errors in the emulator predictions, hence why we see the fresh emulator takes the longest to converge. It is worth noting that greater computational performance improvements could likely be achieved by using an MCMC sampler designed for GPU based sampling, which we do not investigate here, but mention as a possibility for future work.

To fine tune each emulator before it is used to generate the posterior samples, we create a fine-tuning data-set of 100 spectra using a Latin-hypercube sampling of the full prior range. Each emulator is then trained on these samples over 64 epochs. The resulting parameter constraints obtained from each sampling are shown as contour plots in Figure \ref{fig:mcmc_contours}. (Note that here we combine $\Omega_c$ and $\Omega_b$ into the total matter fraction, $\Omega_m$, as weak lensing is primarily sensitive to their sum).

We see that all emulators recover the baseline constraints obtained using CCL reasonably well, though the fresh emulator is somewhat biased. Encouragingly, the MAML emulator appears to perform the best, with constraints well-centred on the CCL-based constraints, although in some parameters does appear to slightly over-constrain. The difference between the MAML and the single-task emulator's constraints appears small. 

\begin{table}
    \centering
    \begin{tabular}{ccc}
        \hline
        Parameter                     & Fiducial value & Prior            \\ \hline
        $\Omega_c$                    & 0.27           & U(0.17, 0.4)      \\
        $\Omega_b$                    & 0.045          & U(0.03, 0.07)    \\
        $h$                           & 0.67           & U(0.4, 1.1)    \\
        $\sigma_8$                    & 0.83           & U(0.65, 1.0)     \\
        $n_s$                         & 0.96           & U(0.8, 1.1)    \\
        $\delta_z^i$                  & 0.00           & U(-0.004, 0.004) \\ \hline
    \end{tabular}
    \caption{Parameter values of the fiducial data vector used in the MCMC analyses and priors on the sampled values. The prior ranges are chosen in order to provide ample space for the posterior distributions to be sampled without being cut off by the prior, except in the case of $\delta_z^i$ where an informative prior is used, whilst also ensuring the emulators are not required to train over a significantly larger region of parameter space than necessary.}
    \label{tab:mcmc_values_priors}
\end{table}

\begin{figure}
    \centering
    \includegraphics[width=0.5\textwidth]{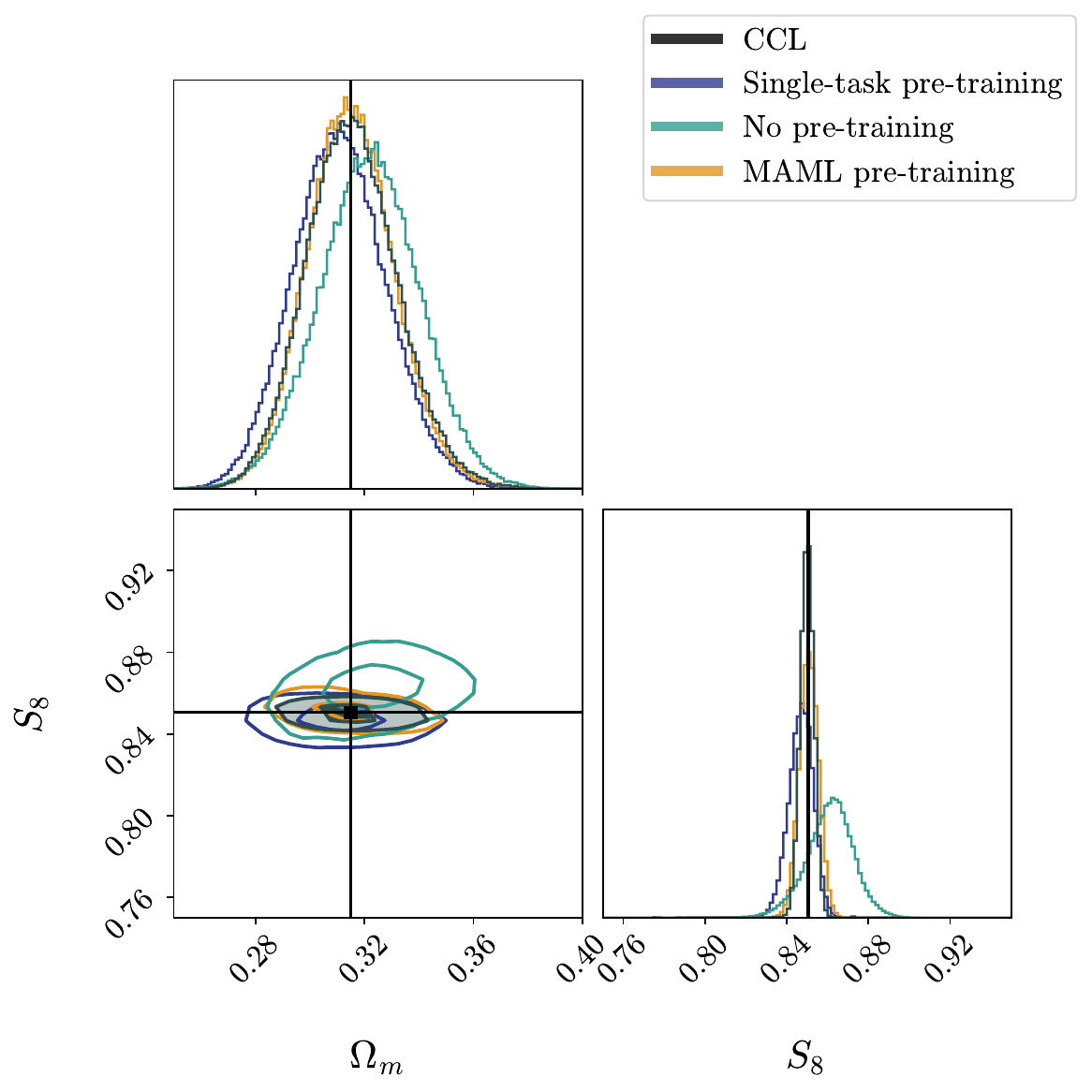}
    \caption{Constraints on $S_8$ and $\Omega_m$ from CCL-based and emulator-based MCMC chains. The same conventions as in Figure \ref{fig:mcmc_contours} are used. In this plane we see a more noticeable difference between the three emulators, with MAML clearly recovering the best constraints.} 
    \label{fig:S8_compare}
\end{figure}

As weak lensing and cosmic shear primarily depend on the total matter fraction, $\Omega_m$, and the amplitude of matter density fluctuations, $\sigma_8$, we calculate the commonly studied parameter $S_8 = \sigma_8\sqrt{\Omega_m/0.3}$ and visualise the constraints in the $S_8$ -- $\Omega_m$ plane in Figure \ref{fig:S8_compare}. This allows us to compare the performance of each emulator in greater detail, focusing on the parameters to which the emulated spectra are most sensitive.

In this plane, we see the MAML emulator clearly obtains the most similar constraints to the baseline constraints. This suggests that the similarity seen between the emulator's constraints on the parameters to which cosmic shear is less sensitive, in Figure \ref{fig:mcmc_contours}, may be partly due to the covariance modelled for the fiducial data-vector and the weaker dependence of the cosmic shear angular power spectrum on these parameters. This is encouraging for the use of MAML, as it shows the small performance gains observed in Section \ref{sec:maml_vs_single} do translate into more accurate constraints on the most sensitive parameters. 

However, it is important to note that the bias to inferred parameter values as a result of using the emulators in place of the Boltzmann codes used by CCL is considerably lower in all cases than potential biases which could arise due to the misspecification of model systematics, the later of which could realistically reach $1-2\sigma$ in a cosmic shear analysis (see, for example, ~\citealt{Campos2023_IA_bias}).

We also tested chains using different numbers of fine-tuning samples and found beyond 500 samples, all emulators produced very similar results. This is not unexpected, as providing more fine-tuning samples will allow all the emulators to learn better representations of the emulated function. The implication here is therefore that the MAML emulator is able to achieve close to its peak performance (within an MCMC chain and given the covariance matrix we have constructed) after 100 fine-tuning samples, while the other emulators require comparatively more to produce the same results.

In the simplistic scenario we consider here, the additional computation time of 500 fine-tuning samples compared to 100 is not significant. However, for more complex scenarios where the evaluation of each sample is more expensive and the model may require more fine-tuning samples to produce accurate constraints, the lower fine-tuning data requirement of MAML may be more desirable, if the pre-training process does not pose a problem.

To provide a quantitative metric of how the constraints from each emulator compare to the baseline, we calculate the Bhattacharyya distance, $D_B$, for the 2D $S_8 - \Omega_m$ emulated posterior distributions and the baseline posterior distribution. In the Gaussian case, the Bhattacharyya distance~\citep{Battacharyya1946} is defined as,
\begin{align}
D_B(P, Q) =  \frac{1}{8} (\mu_P - \mu_Q)^T \Sigma^{-1} (\mu_P - \mu_Q) 
 + \frac{1}{2} \ln \left( \frac{|\Sigma|}{\sqrt{|\Sigma_P| |\Sigma_Q|}} \right),
\end{align}
where $\Sigma = \frac{1}{2}(\Sigma_P + \Sigma_Q)$. $\mu_p$ and $\mu_q$ denote the mean vectors of distributions $P$ and $Q$, with $\Sigma_p$ and $\Sigma_q$ as their covariance matrices. We take the baseline posterior samples as distribution $P$, and the posterior for each emulator as distribution $Q$. For the MAML emulator, we find $D_B = 0.008$, for the single-task pre-trained emulator, $D_B = 0.038$, and for the freshly trained emulator, $D_B = 0.243$. This reinforces the previous observation that the posterior distribution obtained using the MAML emulator is indeed the most similar to the baseline posterior.

\section{Conclusions}
\label{sec:conc}

In this work, we have investigated the use of the MAML training algorithm in order to train a cosmic shear angular power spectrum emulator for rapid adaptation to different galaxy samples, with very few training data. Whilst MAML has been extensively studied in the field of machine learning, its use for training cosmological emulators has not until now been explored. As such, our focus was on a controlled toy problem with a simplified treatment of survey uncertainties, which allowed us to investigate in detail MAML's ability to adapt to changing source redshift distributions, including the computational requirements and performance of MAML for cosmological emulation as compared to standard training methods both in terms of raw emulation accuracy and MCMC derived cosmological constraints.

Future work should seek to determine whether the results found here extend to cases where the emulated spectra and task diversity are more complex still. A particularly natural extension would be to explore MAML's ability to adapt to changing systematic models, and to further motivate this we included in Appendix \ref{sec:ia} an exploratory investigation of MAML in a highly simplified intrinsic alignment scenario. Additional extensions could involve the inclusion of baryonic feedback, changing gravity models and dark energy parametrisations, and more realistic treatments of photometric redshift uncertainty.

We also devised a neural network architecture for power spectrum emulation based on convolutional layers, with the goal of better capturing correlations between different tomographic bin combinations in the emulated data-vector. Whilst we did not investigate in detail the optimal architecture here, it would be interesting to consider in future whether certain architectures perform better with MAML training, particularly for cosmological emulation.

Our grid search analysis to identify the optimal number of different training tasks (unique redshift distributions), samples (spectra generated using different sets of cosmological parameters) and the number of tasks to include in each training batch, showed diminishing returns when the total number of training spectra (the product of the number of tasks and samples) exceeded 10,000. We therefore settled on training using 20 tasks, 500 samples per task, and 5 tasks per batch for the best balance between model performance and the volume of training data required. This relatively low number of tasks suggests the task complexity and/or the variation in the network required to produce accurate results for each task is low, further motivating the need to increase the complexity of the emulation problem in future work, in order to test more rigorously the potential benefits of MAML training.

When comparing the performance of the MAML and single-task pre-trained emulators on a novel unseen task in the context of changing redshift distributions, we found the MAML emulator produced on average more accurate emulations and showed less variability in its performance when provided with different training samples. This indicates that in the scenario considered here, MAML training does provide measurable performance gains when fine-tuning an existing emulator to a novel task. The additional pre-training overhead of MAML is negligible in the case where a GPU is available, though becomes more significant if the emulators are trained using CPUs. It is likely the additional overhead will become more significant as task complexity increases and a larger volume of pre-training data and tasks is required. Future work should seek to identify the scaling of this overhead relative to the performance gains observed with MAML.

Similarly, attempting to train an emulator from scratch for a new problem in this case required significantly more data to achieve the same performance as the fine-tuned MAML emulator. If we also consider the pre-training data used for MAML, the freshly trained emulator required about 80\% of the number of samples used to pre-train and fine-tune the the MAML emulator. This suggests in cases where an emulator is likely to be re-purposed / re-used MAML training is highly preferable to building a new emulator from scratch, though again, it remains the subject of future work to determine whether this holds for more complex emulation scenarios.

In order to place our results in the context of cosmological inference, we investigated the use of all three emulators to generate posterior samples within an MCMC analysis. Use of the MAML emulator produced the most similar posterior distributions to a baseline comparison calculated from theory, whilst all emulators were able to recover the majority of the baseline constraints with reasonable accuracy. For the most sensitive cosmic shear parameters of $S_8$ and $\Omega_m$, we saw a larger discrepancy in the emulator constraints, with MAML clearly producing the most accurate constraints relative to the baseline. Future work should consider the impact of sampling choices on these constraints, such as the use of different prior ranges, walker initialisations, or even sampling algorithms. The consideration of additional systematics would also lead to larger uncertainties and potentially more similar results between each emulator.

Our results indicate that MAML shows promise as a means of training adaptable cosmological emulators. In future, such an approach may enable the development of more generic power spectrum emulators, capable of adapting not only to novel galaxy samples, but also to other classes of modelling uncertainty, including intrinsic alignment, baryonic feedback, alternative gravity theories, and dark energy parametrisations. Future work is essential to determine whether these extensions are feasible in practice, whether the potential performance gains justify the additional training cost, and how those gains scale with increasing task complexity.

\section*{Acknowledgements}

We thank Marika Asgari for helpful discussions. This work used in part the computational facilities DiRAC@Durham and DiRAC@Cambridge, managed by the STFC DiRAC HPC Facility (\url{www.dirac.ac.uk}). CMG is supported by the Newcastle University Lady Bertha Jeffreys studentship. CDL is supported by the Science and Technology Facilities Council (STFC) [grant No. UKRI1172]. This work would not have been possible without the Python libraries \textsc{SciPy}~\citep{2020SciPy-NMeth}, \textsc{PyCCL}~\citep{Chisari_2019}, \textsc{Numpy}~\citep{Numpy}, \textsc{PyTorch}~\citep{paszke2019pytorchimperativestylehighperformance}, and \textsc{TJPCov} (\url{(https://github.com/LSSTDESC/TJPCov}). This result is part of a project that has received funding from the European Research Council (ERC) under the European Union's Horizon 2020 research and innovation programme (Grant agreement No. 948764; PB). For the purpose of open access, the author has applied a Creative Commons Attribution (CC BY) licence to any Author Accepted Manuscript version arising from this submission.

Author contributions: CMG led the direction of the project and performed the majority of analysis (writing and validating code, writing this paper). CDL helped shape the direction of the project and provided supervisory support and guidance on key cosmological concepts, as well as contributions to the text. MR provided ideas and guidance related to machine learning and statistics. PB provided the initial motivation for this work and helped formulate the direction of the work.

\balance

\section*{Data Availability}

All data used in this work has been produced by the authors. Source code used to produce this data, as well as the results presented can be found in the associated GitHub repository found at \url{https://github.com/CMacM/CosyMAML}. In cases where the data used is too large to store on GitHub, the authors can provide it to interested parties upon reasonable request.

\balance

\section*{Conflict of Interest}

The authors declare that they have no conflicts of interest.



\bibliographystyle{rasti.bst}
\bibliography{example.bib} 




\appendix

\section{Computational Requirements of Power Spectrum Projection}
\label{app:proj}

When calculating from theory APS within a cosmological inference pipeline, there are two distinct stages to the computation. Initially, the sampled cosmological parameters must be used to construct the 3D matter power spectrum, $P_{\delta}$, which does not inherently depend on the galaxy sample in question. To obtain the APS, $P_{\delta}$ must be projected via a Limber integral (Equation \ref{eqn:lensing_aps}) over the lensing/clustering kernels, which themselves depend on the redshift distribution of the galaxy sample (Equation \ref{eqn:lensing_kernel}).

Computation of $P_{\delta}$ is a challenging process, requiring the use of Boltzmann codes such as CLASS~\citep{Lesgourgues2011boltzmannclass}, whilst the projection over the lensing/clustering kernels is a relatively simple integration. However, it is important to consider that although the projection for a single angular power spectrum is computationally inexpensive, likelihood evaluations in MCMC analyses require many such projections, resulting in a non-negligible cumulative cost. 

For example, real surveys often use upwards of five distinct tomographic bins for their galaxy samples and seek to compute APS from the cross-correlation of multiple redshift bins. For $5$ bins, this leads to $15$ unique combinations. In a 3x2pt analysis, where the cosmic shear (lensing-lensing) galaxy-galaxy lensing (lensing-clustering) and clustering APS are required, the number of required projections further increases. For example, in the Dark Energy Survey Year 6 results (\citealp{DES_Y6_Results}; DES Y6), $45$ unique projections were required for their 3x2pt analysis, derived from five lens bins (originally six, but one was excluded due to data issues), and four source bins. Based on the LSST science requirement document~\citep{LSST_SRD}, the Year 10 data vector will contain up to $75$ unique projections. As such, for every set of cosmological parameters sampled in an MCMC analysis, $P_{\delta}$ need only be calculated once, but projection must be performed many times.

In order to understand the potential benefits of emulating the angular power spectrum directly, as opposed to only emulating the matter power spectrum, we time each step in the APS computation. CCL allows us to compute the matter power spectrum separately, before then passing the pre-computed matter power spectrum for projection. For our tests, we sample $100$ unique sets of cosmological parameters and for each set, we then compute the matter power spectrum once and time this step. We then loop over the projection step $15$, $45$, and $75$ times using the pre-computed matter power spectrum, and measure the time to complete the projection loop. These values correspond to five bins cosmic shear only, the DES Y6 3x2pt analysis projection count, and an LSST Year 10-like analysis. It is important to note that to mimic what occurs in Section \ref{sec:mcmc} where individual walkers are run in parallel on separate cores, we force matter power spectrum computation and projection to run on a single CPU core. The results are shown in Table \ref{tab:proj_compute}

\begin{table}
\centering
\begin{tabular}{l|ll}
\hline
Analysis type         & Time for projection & Time for $P_{\delta}$ \\ \hline
5 bins, cosmic shear  & 6.94\%                & 93.1\%             \\
DES Y6, 3x2pt & 18.6\%              & 81.4\%             \\
LSST Y10, 3x2pt        & 27.3\%               & 72.7\% \\            
\hline
\end{tabular}
\caption{Results of timing analysis on relative compute time required for matter power spectrum computation and projection into angular power spectra. Although the projection time is small, because projection must be done multiple times for each set of cosmological parameters, its effect on the total runtime is cumulative. For a LSST Year 10-like 3x2pt analysis with ten unique tomographic bins, projection alone can comprise a substantial fraction of the total compute time.}
\label{tab:proj_compute}
\end{table}

We emphasise that the values we present here are intended purely as a guide to help with interpretation of the motivation behind this work and the results presented in the main body of the paper. We do not perform detailed analysis of how these values might change for different MCMC configuration, or how shared usage on HPC systems could impact these values. As such, they are not precise estimates, but rather a quantitative guide to help understand the distribution of computational burden in a cosmological inference pipeline. Nonetheless, these results indicate that emulating only the matter power spectrum does not eliminate all significant computational cost, as projection remains a non-negligible component of the likelihood evaluation.

\section{Intrinsic Alignment as a Meta-Learning Task}
\label{sec:ia}

In weak lensing, a key systematic that must be modelled is the intrinsic alignment (IA) of galaxies, which contaminates weak lensing measurements by inducing correlations in galaxy shapes similar to those used to measure the lensing signal. IA remains an active field of research, with many competing models (for a summary of contemporary IA research and models, see~\citealp{Lamman_2024}), and it has been shown that adopting an incorrect IA prescription can significantly bias cosmological inference~\citep{Secco_2022}.

As discussed in Section~\ref{sec:intro}, a potential future application of MAML is to provide flexibility to different systematic models within weak lensing analyses, enabling researchers to explore alternative modelling assumptions while retaining the benefits of emulator-accelerated inference. Additionally, IA represents a meaningfully more challenging source of variation between tasks than changes in the source redshift distribution, as it alters the structure of the cosmic shear APS in a scale- and redshift-dependent manner.

Motivated by this, we conduct an exploratory analysis to address the question: can MAML be used to implicitly learn the impact of intrinsic alignment contamination on cosmic shear angular power spectra? The analysis of this section is not intended to represent a realistic survey pipeline. In particular, we do not consider multiple IA models, leaving this to future work. Rather, our aim is to test whether MAML can adapt to changing IA contaminations that originate from varying the parameter values within a single IA model, without the emulator receiving any explicit information about IA as inputs.

To this end, we construct a distribution of meta-learning tasks in which different intrinsic alignment contaminations define different tasks. These contaminations are generated using a redshift-dependent non-linear alignment (NLA) model~\citep{Bridle_2007,Krause2016}, with different choices of NLA parameter values corresponding to different tasks in the MAML framework.

In the NLA model, the intrinsic alignment contribution is typically given in terms of a redshift-dependent amplitude. We present this amplitude following the convention of~\cite{Secco_2022},
\begin{equation}
A_{\rm IA}(z) = -a_1\,\bar{C}\,\frac{\rho_{\rm crit}\,\Omega_m}{D(z)}\left(\frac{1+z}{1+z_0}\right)^\eta,
\end{equation}
where $\bar{C}=5 \times 10^{-14} \ M_{\odot} \ h^{-2} \ \mathrm{Mpc}^2$ is a normalisation constant, $\rho_{\rm crit}$ is the critical density of the Universe today, $\Omega_m = \Omega_c + \Omega_b$ as mentioned above, and $D(z)$ is the linear growth factor of matter. Together, the aforementioned parameters define the standard normalisation of the NLA model, while $a_1$ sets the overall alignment amplitude and $\eta$ controls the redshift scaling. The pivot redshift $z_0$ is typically fixed for a given survey, we follow \cite{Secco_2022} and fix its value to $0.62$.

We vary the NLA parameters $a_1$ and $\eta$ across tasks. Crucially, these IA parameter values are not provided as inputs to the emulator, and the MAML model must therefore implicitly learn to adapt to their impact on the cosmic shear angular power spectra.

For testing, the IA parameters are fixed to $a_1 = 0.175$ and $\eta = 3.90$, and the MAML model is fine-tuned to angular power spectra generated with this fixed IA prescription. Inference is then performed on the cosmological parameters only. To isolate the question of how well MAML can handle IA and keep this distinct from our primary study of changing redshift distributions, we fix the redshift distribution to the LSST Y1 distribution and do not include shifts in the tomographic bin means.

We emphasise that in realistic survey analyses, IA parameters are typically jointly constrained or marginalised over. Given that our implementation of the NLA model has only two varying parameters, these could straightforwardly be included as emulator inputs in a standard approach. However, our objective here is instead to assess whether MAML can implicitly learn to adapt to changing IA contaminations, as a first step towards future applications involving adaptation to fundamentally different IA models, rather than to mimic a realistic analysis.

To train the MAML model, we uniformly sample $20$ sets of NLA parameters from the ranges $a\in(0, 0.79)$ and $\eta\in(0.61, 4.92)$. These ranges are chosen based on the 1-sigma NLA constraints found in~\citealp{Secco_2022}. For each set of NLA parameters, we then generate a 5-dimensional hypercube of $500$ cosmological parameter samples using the same ranges given in Table \ref{tab:params}. Our MAML training uses the same set-up as used in Section \ref{sec:building_training}, with two modifications: we do not use a stopping criteria based on a validation set (instead training for a fixed number of $200$ epochs), and we lower the initial outer learning rate to $\mathcal{A} = 0.05$. The latter choice was motivated by large fluctuations in the per-epoch training loss observed at higher learning rates, indicative of overfitting to individual meta-task batches.

We run a baseline MCMC analysis with the aforementioned fixed IA parameter values, using CCL in the same configuration as presented in Section \ref{sec:mcmc}, with one modification: we run the chains until the total chain length exceeds $100$ times the autocorrelation time for all parameters. This is done to give more samples and thus smoother contours. In this case the, chain converged after $9,000$ steps, equivalent to approximately $725$ CPU core-hours.

The MAML-based MCMC reached convergence in $10,000$ steps, equivalent to approximately $167$ CPU core-hours or $0.5$ GPU hours. For both the CCL and MAML chains, convergence is significantly faster than in Section \ref{sec:mcmc} in this case because of the lower parameter space dimensionality. We report these timings only for completeness; the focus of this section is the representational capability of MAML rather than computational efficiency.

\begin{figure}
    \centering
    \includegraphics[width=0.5\textwidth]{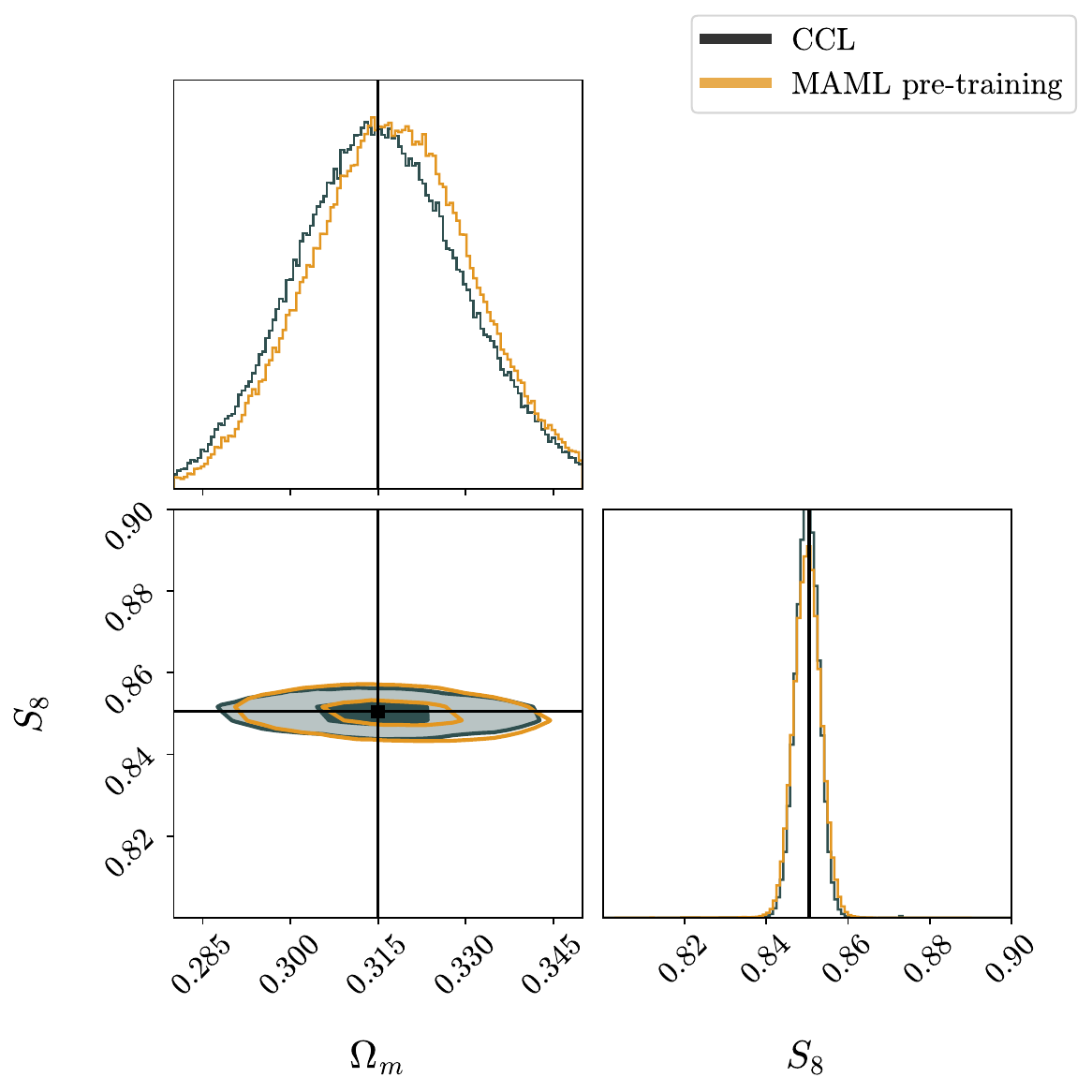}
    \caption{Constraints on $S_8$ and $\Omega_m$ from CCL-based and MAML emulator-based MCMC chains, using MAML training to adapt to new set of NLA parameters. Although we observe a very slight shift in the constraints from the MAML emulator, both sets of contours largely overlap and capture the fiducial values within the 1-sigma contours.}
    \label{fig:s8-Om-IA}
\end{figure}

We present the results of this analysis as constraints on $S_8$ and $\Omega_m$ in Figure \ref{fig:s8-Om-IA}. We see very good recovery of the CCL-based posterior by the MAML emulated chain with both chains successfully capturing the fiducial values within the 1-sigma confidence intervals. We also compute the Battacharyya distance between the MAML and CCL contours and find a value of $D_{B} = 0.007$. Our results indicate that MAML can enable adaptation to changing IA contaminations derived from the same IA model in this controlled setting. Future work should seek to extend this analysis to include changing the underlying IA model to distinct prescriptions, such as the halo model~\citep{Fortuna_2020} or hybrid Lagrangian model~\citep{Maion_2024} among others.


\bsp	
\label{lastpage}
\end{document}